\documentclass[aps,11pt,preprintnumbers,nofootinbib,floatfix]{revtex4}

\usepackage{graphicx}% Include figure files
\usepackage{epsfig}
\usepackage{dcolumn}% Align table columns on decimal point
\usepackage{subfigure}
\usepackage{multirow}
\usepackage{amsmath}
\usepackage{amssymb}

\begin{document}

\title{Heat engine efficiency and Joule-Thomson expansion of non-linear charged AdS black hole in massive gravity}

%%%% To generate auto affiliation numbers please use \author{}\affil{} command

\author{Cao H. Nam}
\email{hncao@yonsei.ac.kr}\affiliation{\small High Energy Physics and Cosmology Laboratory, Phenikaa Uni‎versity, Hanoi 100000, Vietnam}
\affiliation{Department of Physics, College of Science, Yonsei University, Seoul 120-749, Korea}
\date{\today}

\begin{abstract}%
In this paper, we have considered the heat engine and Joule-Thomson expansion for the charged AdS black hole in the context of the non-linear electrodynamics and massive gravity. For the black hole heat engine, we obtained the analytical expression for the efficiency in terms of either the entropies or the temperatures and pressures in various limits. For the Joule-Thomson expansion of the black hole, we derived the isenthalpic curves in $T-P$ diagram, the Joule-Thomson coefficient, and the inversion curves. We also indicated in detail the effects of the non-linear electrodynamics and massive gravity on the heat engine efficiency and the Joule-Thomson expansion of the black hole.
\end{abstract}

\maketitle

\section{Introduction}
Within the framework of General Relativity (GR), under general physical conditions, the spacetime would admit curvature singularities, i.e., the boundaries of spacetime beyond which an extension of spacetime is impossible and at which the curvatures and densities become infinite \cite{Hawking1973}. In the cosmology, the beginning of Universe was at big bang singularity. The gravitational collapse of the matters leads to the formation of the black holes with the curvature singularities surrounded by the event horizon. It is widely believed that the presence of these curvature singularities are a sign indicating the breakdown of GR and thus requiring a more fundamental theory of the gravitation, e.g. quantum gravity. But, such a complete theory has not been achieved so far. Since, whether the spacetime singularity can be resolved at the level of the classical gravity is still an open question in current research.

The non-linear electrodynamics may appear as the low-energy limit of the heterotic string theory 
\cite{Natsuume1994,Padi2007,Sun2008,Szepietowski2009,Pastras2009}. On the other hand, the classical gravity with the nonlinear electrodynamics can be realized as the effective description arising from quantum gravity coupled to the matter. Thus, it is natural to expect that the non-linear properties of the fundamental theory may be exhibited in the physics of the black holes. In particular, Ayon-Beato and Garcia has indicated that a source of the non-linear electrodynamics causes a regular black hole which is free of a curvature singularity at the origin but possesses an event horizon \cite{Ayon-Beato98,Beato1999}. In addition, the Bardeen black hole, which is the first regular black hole found by Bardeen \cite{Bardeen} but whose physical source was not realized at that moment, was later reobtained as a gravitational collapse of some magnetic monopole in the context of the non-linear electrodynamics \cite{Ayon-Beato2000}. Since the pioneering works of Ayon-Beato and Garcia, many regular black hole solutions with various non-linear electromagnetic sources have been found in the literature  \cite{Cataldo2000,Bronnikov2001,Burinskii2002,Dymnikova2004,Matyjasek2004,Berej2006,Fabris2006,Balart2014,
Houndjo2015,Fan-2016,Rodrigues2016,Marques20016,Sajadi2017,Toshmatov2017,Singh2018,Nam2018a,Ali-Ghost2018,Nam2018b,Silva2018,Nam2018d,Kumar2018}. Other regular black holes were also constructed \cite{Dymnikova1992,Borde1994,Frolov1996,Cabo1999,Nicolini2006,Hayward2006,Bambi2013}. Studying the regular black holes has thus drawn many attentions
\cite{Kim-Park2009,Gonzalez2009,Myung2009,Li-Lin2013,Modesto2013,Aftergood2014,Ma-Zhao2014,Toshmatov,Hendi2015c,Huang2015,Hendi2015d,Junior,Ghosh,Lin-Yang2015,Zangeneh2015,Gan2016,Dehghani2017,Nojiri2017,Rincon2107,Shahzad2017,Leon2017,CWu2018,
Maluf2018,
Jusufi2018,Kofane2018,Bambi2018,YH-Wei2018,Nam2018c,Roman2018,Rubio2018,Sharif2018,Javed2018,Hu2019,Zhang2018,Jawad2018,Cano2019,Schee2019,
Estrada2019}.

One of the straightforward extensions of GR is to consider that graviton is massive spin-$2$ particle with five degrees of freedom. A theoretical reason for this extension is that it could
provide the natural resolution for the acceleration of Universe without introducing dark energy.
Also, the recent observations of the gravitational waves by LIGO have constrained graviton mass to $m\leq1.2\times10^{-22}$ eV \cite{LIGO}. This means that the mass of graviton can be tiny but non-zero.
The first construction for the massive gravity theory, which is ghost-free, was done by Fierz and Pauli (FP) \cite{Fierz1939}, at which mass term at the inearized level is included as
\begin{equation}
 S_{\text{FP}}=-\frac{m^2}{4}\int d^4x\left(h_{\mu\nu}h^{\mu\nu}-h^2\right),
\end{equation}
where $h_{\mu\nu}$ is a symmetric tensor field describing massive graviton, $h^{\mu\nu}=\eta^{\mu\sigma}\eta^{\nu\rho}h_{\sigma\rho}$, and $h=\eta^{\mu\nu}h_{\mu\nu}$. However, this massive gravity theory suffers from an important problem, which GR is not recovered in the zero mass limit of graviton, well-known as the van Dam-Veltman-Zakharov (vDVZ) discontinuity \cite{Dam1970,Zakharov1970}. The vDVZ discontinuity could be resolved 
in the nonlinear massive gravity theories which include additionally the higher derivative terms \cite{Vainshtein1972,Boulware1972a,Boulware1972b}. Unfortunately, these non-linear massive gravity theories lead to the appearance of the so-called Boulware and Deser (BD) ghosts. A successful non-linear massive gravity theory, which is ghost-free and recovers GR as graviton mass goes to zero, was proposed by
de Rham, Gabadadze and Tolley (dRGT) \cite{deRham2010,deRham2011}. Later, various black hole solutions have been found in dRGT massive theory and their thermodynamics as well as critical phenomena have been investigated
\cite{Gao2013,Komada2014,Cai2015,Gosh2016,Panahiyan2016a,Panahiyan2016b,Prasia2016,
Panahiyan2016c,Ning2016,Panah2017,Zou2017,Tannukij2017,Guo2017,
Yue2017,Boonserm2018,Nam2018d}.

The study of the black hole thermodynamics in anti-de Sitter (AdS) space has been an interesting research topic because of its rich phase structure as well as AdS/CFT correspondence \cite{Maldacena}. Hawking and Page
discovered a phase transition between Schwarzschild AdS black hole and thermal AdS space \cite{Hawking1983}, called the Hawking-Page transition in the literature, which is interpreted as the confinement/deconfinement phase transition in the boundary conformal field theory \cite{Witten,Gubser}. In addition, Chamblin \textit{et al.} studied the charged AdS black holes in both canonical and grand canonical ensembles, and they found a first-order phase transition between small and large black holes \cite{Chamblin199a,Chamblin199b}.
Recently, the phase space of the AdS black hole thermodynamics has been extended at which the cosmological constant is treated as thermodynamic pressure corresponding to the conjugate quantity as the thermodynamic volume \cite{Wang2006,Kastor2009,Kastor2010,Dolan2011,Dolan2011b,Teo2017}. A a result, the black hole mass is most naturally considered as the enthalpy, rather than the internal energy. In the extended phase space, $P-V$ criticality was discovered in the charged AdS black hole \cite{Mann2012} as well as in various black holes
 \cite{Hendi2012,Cai2013,Liu2013,MoLiu2014,Liu2014,Li2014,Zhao2014,Dehghani2014,Hennigar2015,Xu2015,
Talezadeh2016,Sadeghi2016a,Liang2016,Fernando2016,Fan2016,Sadeghi2016b,Hansen2017,Majhi2017,Talezadeh2017, Upadhyay2017,Pradhan2018,Aydner-2019}.

With the extended phase space, it is possible to extract the mechanical work from the heat
energy via the term $PdV$. This suggests that the concept of the traditional heat engine can be incorporated into the black holes at which the black holes play the role as the working substances \cite{Johson2014,Belhaj2015,Setare2015,Johson2016a,Johson2016b,Zhang2016,Bhamidipati2017,Hennigar17,
Mo2017,Hendi2018,Chakraborty2018,Ghaffarnejad2018,FangKuang2018,ZhangYu2018,Rosso2018,MoLan2018,Panah2018,
Santos2018,Lobo2018,Johnson2019,HuKuang2019,Fernando2019,ZhangYu2019,Yaraie2019,Debnath2019,Rajani2019}. A heat engine is defined by a heat cycle which is a closed path in the $P-V$ diagram. It works between warmer and colder reservoirs which correspond to the temperatures $T_H$ and $T_C$ ($T_H>T_C$), respectively. During the working process, the heat engine absorbs a heat amount $Q_H$ from the warmer reservoir and exhausts a heat amount $Q_C$ to the colder reservoir. A total mechanical work $W$, which is produced by the heat engine, is given by $W=Q_H-Q_C$. The efficiency of the heat engine is defined by
\begin{equation}
\eta=\frac{W}{Q_H}=1-\frac{Q_C}{Q_H}.
\end{equation}
It is clear that the efficiency of the black hole heat engine depends crucially on both the equation of the state provided by the black hole and the the paths forming the heat cycle in the $P-V$ diagram. On the other hand, various black holes should produce the different heat engines.

One of the recent developments with respect to the black hole thermodynamics in the extended phase space is the Joule-Thomson expansion. In the traditional thermodynamics, the Joule-Thomson expansion is that the enthalpy is constant during the expansion process of the gas or fluid from a high pressure to a low pressure through a porous plug. Intrestingly, \"{O}kc\"{u} and Aydiner incorporated the concept of the Joule-Thomson expansion into the charged AdS black holes \cite{Aydner2017a} and the Kerr-AdS black holes \cite{Aydner2017b}. Later, the Joule-Thomson expansion has been generalized to various black holes \cite{Haldar2018,Mo-Li2018,QLan2018,Chabab2018,MoLi2018,Cisterna2018,LiHe2019,Kumara2019,Guo2019,PuGuo2019,Yekta2019}.

This work is organized as follows. In Sec. \ref{bhsol}, we review briefly the non-linear charged AdS black hole solution and its thermodynamics in the massive gravity, obtained in Ref. \cite{Nam2018d}. In Sec. \ref{heat-engine}, we treat this black hole as the heat engine. In this section, we compute the efficiency of the black hole heat engine, study how the non-linear parameter as well as the massive gravity couplings affect the efficiency of the black hole heat engine, and then compare it to the Carnot efficiency. Sec. \ref{JT-expansion} is devoted to study the  Joule-Thomson expansion of the non-linear charged AdS black hole in the massive gravity. Finally, we make a conclusion in last section. Note that, in this paper, we use units in $G_N=\hbar=c=k_B=1$ and the signature of the metric $(-,+,+,+)$.

\section{\label{bhsol} A brief review of non-linear charged A$\textrm{d}$S black hole}

The aim in this present work is to generalize the recent developments of the heat engine and Joule-Thomson expansion for the non-linear charged AdS black hole in massive gravity, which was recently obtained in Ref. \cite{Nam2018d}. Thus, in this section we will review this black hole solution and its thermodynamic properties. This black hole solution obtained from solving the equations of motion for the system of the massive gravity coupled to the non-linear electromagnetic field in the four-dimensional AdS spacetime background. The action of this system is given by
\begin{equation}
S=\int
d^4x\sqrt{-g}\left\{\frac{1}{16\pi}\left[R-2\Lambda+m^2\sum^4_{i=1}c_i\mathcal{U}_i(g,f)\right]-\frac{1}{4\pi}\mathcal{L}(F)\right\},\label{EM-nlED-adS}
\end{equation}
where $R$ refers to the spacetime scalar curvature, $\Lambda$ is the negative cosmological constant expressed in terms of the curvature radius $l$ of
the AdS spacetime background as
\begin{equation}
\Lambda=-\frac{3}{l^2},
\end{equation}
$m$ is graviton mass, $c_i$ are the couplings of the massive gravity,
$f$ is the reference metric which is kept fixed, $\mathcal{U}_i$ are symmetric polynomials in terms of the
eigenvalues of the $4\times4$ matrix
${\mathcal{K}^\mu}_\nu=\sqrt{g^{\mu\lambda}f_{\lambda\nu}}$ given
as
\begin{eqnarray}
\mathcal{U}_1&=&[\mathcal{K}],\nonumber \\
\mathcal{U}_2&=&[\mathcal{K}]^2-[\mathcal{K}^2],\nonumber \\
\mathcal{U}_3&=&[\mathcal{K}]^3-3[\mathcal{K}][\mathcal{K}^2]+2[\mathcal{K}^3],\nonumber\\
\mathcal{U}_4&=&[\mathcal{K}]^4-6[\mathcal{K}]^2[\mathcal{K}^2]+8[\mathcal{K}][\mathcal{K}^3]+3[\mathcal{K}^2]^2-6[\mathcal{K}^4],
\end{eqnarray}
with $[\mathcal{K}]={\mathcal{K}^\mu}_\mu$. The Lagrangian $\mathcal{L}(F)$ of the non-linear electrodynamics $\mathcal{L}(F)$ is defined as
\begin{equation}
\mathcal{L}(F)=Fe^{-\frac{k}{2Q}\left(2Q^2F\right)^{\frac{1}{4}}},\label{nl-Lag} \ \ F\equiv\frac{1}{4}F_{\mu\nu}F^{\mu\nu},
\end{equation}
where $F_{\mu\nu}=\partial_\mu A_\nu-\partial_\nu A_\mu$ is the strength tensor of the non-linear electromagnetic field, $Q$ is the total charge of the system, and $k$ is a fixed parameter by which the charge $Q$ and the mass $M$ of the system are related as, $Q^2=Mk$.

With the spherically-symmetric and static spacetime and the reference metric taken in the following form \cite{Cai2015}
\begin{equation}
f_{\mu\nu}=\textrm{diag}(0,0,c^2,c^2\sin^2\theta),
\end{equation}
where $c$ is a positive constant, we find the equations of motion which are obtained from the variation of the action (\ref{EM-nlED-adS}) as
\begin{eqnarray}
G^\nu_\mu-\left[\frac{3}{l^2}+m^2\left(\frac{cc_1}{r}+\frac{c^2c_2}{r^2}\right)\right]\delta^\nu_\mu&=&2\left[\frac{\partial\mathcal{L}(F)}{\partial
F}F_{\mu\rho}F^{\nu\rho}-\delta^\nu_\mu\mathcal{L}(F)\right],\label{Eeq}\nonumber\\
\nabla_\mu\left(\frac{\partial\mathcal{L}(F)}{\partial
F}F^{\nu\mu}\right)&=&0.\label{Meq1}\nonumber\\
\nabla_\mu*F^{\nu\mu}&=&0.\label{Meq2}
\end{eqnarray}
A spherically-symmetric and static black hole solution of the mass $M$ and magnetic charge $Q$ is obtained from solving these equations of motion, given by
\begin{eqnarray}
ds^2&=&-f(r)dt^2+f(r)^{-1}dr^2+r^2d\Omega^2_2,\nonumber\\
F_{\mu\nu}&=&\left(\delta^\theta_\mu\delta^\varphi_\nu-\delta^\theta_\nu\delta^\varphi_\mu\right)B(r,\theta),
\end{eqnarray}
where
\begin{eqnarray}
f(r)&=&1-\frac{2M}{r}e^{-\frac{k}{2r}}+\frac{r^2}{l^2}+m^2\left(\frac{cc_1r}{2}+c^2c_2\right),\nonumber\\
B(r,\theta)&=&Q\sin\theta.
\end{eqnarray}
It is important to note that $1+m^2c^2c_2$ plays the role of the effective horizon curvature which can be positive (the sphere effective horizon), zero (the flat one), or negative (the hyperbolic one) depending on the sign and the absolute value of the coupling parameter $c_2$. With $M=Q=0$, it leads to the vacuum solution as
\begin{equation}
  f(r)=1+\frac{r^2}{l^2}+m^2\left(\frac{cc_1r}{2}+c^2c_2\right).\label{vac-sol}
\end{equation}
In the region of the large distances $k/r\ll1$, the function $f(r)$ becomes
\begin{equation}
  f(r)\simeq1-\frac{2M}{r}+\frac{Q^2}{r^2}+\frac{r^2}{l^2}+m^2\left(\frac{cc_1r}{2}+c^2c_2\right).
\end{equation}
This means that the black hole behaves asymptotically like the $4D$ RN-AdS black hole in the massive gravity \cite{Cai2015,Gosh2016,Panahiyan2016a,Panahiyan2016b}.

The black hole mass $M$ is expressed in terms of the event horizon radius $r_+$ and the pressure $P=-\frac{\Lambda}{8\pi}=\frac{3}{8\pi l^2}$ as
\begin{equation}\label{enthalpy}
  M=\frac{r_+}{2}e^{\frac{k}{2r_+}}\left[1+\frac{8\pi Pr^2_+}{3}+m^2\left(\frac{cc_1r_+}{2}+c^2c_2\right)\right].
\end{equation}
 The first law of the black hole thermodynamics is given by
\begin{equation}\label{1stlaw}
  dM=TdS+VdP+\mathcal{C}_1dc_1+\mathcal{C}_2dc_2,
\end{equation}
where, because in general the massive gravity couplings can vary, they have been treated as the thermodynamic variables corresponding to the conjugating variables $\mathcal{C}_1$ and $\mathcal{C}_2$, respectively.
The black hole temperature is identified by using the surface gravity at the event horizon as
\begin{equation}\label{bhtemp}
  T=\frac{f'(r_+)}{4\pi}=\left(2r_+-\frac{k}{3}\right)P+\frac{(2r_+-k)}{8\pi r^2_+}\left[1+m^2\left(\frac{cc_1r_+}{2}+c^2c_2\right)\right]+\frac{m^2cc_1}{8\pi}.
\end{equation}
The black hole entropy $S$, the thermodynamic volume $V$, and the conjugating quantities $\mathcal{C}_{1,2}$ are obtained from the first law as
\begin{eqnarray}
S&=&\int\frac{1}{T}\left(\frac{\partial M}{\partial
r_+}\right)dr_+=\pi r^2_+\left(1+\frac{k}{2r_+}\right)e^{\frac{k}{2r_+}}-\frac{\pi k^2}{4}\textrm{Ei}\left(\frac{k}{2r_+}\right),\label{bhentr}\\
V&=&\left(\frac{\partial M}{\partial P}\right)_{S,c_1,c_2}=\frac{4\pi r^3_+}{3}e^{\frac{k}{2r_+}},\label{vol}\\
\mathcal{C}_1&=&\left(\frac{\partial M}{\partial c_1}\right)_{S,P,c_2}=\frac{m^2cr^2_+}{4}e^{\frac{k}{2r_+}},\\
\mathcal{C}_2&=&\left(\frac{\partial M}{\partial c_2}\right)_{S,P,c_1}=\frac{m^2c^2r_+}{2}e^{\frac{k}{2r_+}},
\end{eqnarray}
where the function $\textrm{Ei}(x)$ is given by
\begin{equation}
  \textrm{Ei}(x)=-\int_{-x}^{\infty}\frac{e^{-t}}{t}dt.
\end{equation}
Clearly, the black hole entropy is modified by only the non-linear electrodynamics. In the regime of the large horizon radius $k/r_+\ll1$, the black hole entropy is approximately given by
\begin{equation}
  S=\pi k^2\left[\left(\frac{r_+}{k}\right)^2+\frac{r_+}{k}+\frac{3-2\gamma}{8}-\frac{1}{4}\ln\left(\frac{k}{2r_+}\right)+\mathcal{O}\left(\frac{k}{r_+}\right)\right],
\end{equation}
where $\gamma\approx0.577216$ is Euler's constant. This expansion suggests that the black hole entropy satisfies approximately the area law ($S=\pi r^2_+$) in the regime of the large horizon radius. The heat capacity at constant pressure is
\begin{eqnarray}
C_P&=&T\left(\frac{\partial S}{\partial T}\right)_P=\frac{\partial M}{\partial r_+}\left(\frac{\partial T}{\partial r_+}\right)^{-1},\nonumber\\
&=&\frac{\pi r^2_+}{6}\frac{h_1(r_+)}{h_2(r_+)}e^{\frac{k}{2r_+}},
\end{eqnarray}
where
\begin{eqnarray}
h_1(r_+)&=&16\pi Pr^2_+(6r_+-k)+3m^2cc_1r_+(4r_+-k)+6(1+m^2c^2c_2)(2r_+-k),\nonumber\\
h_2(r_+)&=&8\pi Pr^3_+-\left(1+m^2c^2c_2-\frac{m^2cc_1k}{4}\right)r_++k(1+m^2c^2c_2).
\end{eqnarray}
The equation of state  for the black hole can easily be obtained from Eq. (\ref{bhtemp}) as
\begin{equation}\label{equ-state}
  P=\frac{T}{2r_+-k/3}+\frac{2(1+m^2c^2c_2)(k-2r_+)+m^2cc_1r_+(k-4r_+)}{16\pi r^2_+(2r_+-k/3)},
\end{equation}
where $r_+=r_+(V)$ is understood to be a function of the thermodynamic volume $V$, which is determined by Eq. (\ref{vol}). If the effective horizon curvature $1+m^2c^2c_2$ is positive, a critical point appears when the isotherm in the $P-r_+$ diagram has an inflexion point, determined by
\begin{equation}
\left(\frac{\partial P}{\partial
r_+}\right)_T=\left(\frac{\partial^2P}{\partial r^2_+}\right)_T=0.
\end{equation}
From this, we can obtain the critical radius $r_c$, the critical temperature $T_c$ and the critical pressure $P_c$ as
\begin{eqnarray}
 r_c&=&\frac{6k(1+m^2c^2c_2)}{4(1+m^2c^2c_2)-m^2cc_1k},\nonumber\\
 T_c&=&\frac{13(1+m^2c^2c_2)}{81\pi k}+\frac{37m^2cc_1}{216\pi}+\frac{(m^2cc_1)^2k\left(1+m^2c^2c_2+m^2cc_1k/96\right)}{108\pi(1+m^2c^2c_2)^2},\nonumber\\
 P_c&=&\frac{(1+m^2c^2c_2-m^2cc_1k/4)^3}{54\pi k^2(1+m^2c^2c_2)^2}.
\end{eqnarray}
If the temperature is smaller than the critical one, the black hole can undergo a first-order phase transition between the small black hole and the large black hole, which is analogous to the van der Waals phase transition. This phase transition disappears if the temperature is larger than the critical one.
\section{\label{heat-engine} The non-linear charged AdS black hole as a heat engine}

\begin{figure}[t]
% Requires \usepackage{graphicx}\
 \centering
\begin{tabular}{cc}
\includegraphics[width=0.5 \textwidth]{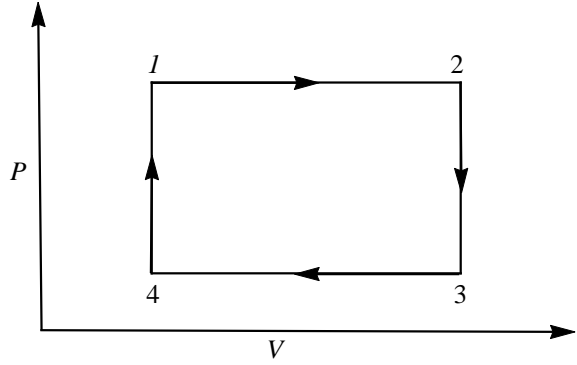}
\end{tabular}
  \caption{A cycle of the black hole heat engine with two isobaric paths and two isochoric paths.}\label{PV-dia}
\end{figure}

We consider a heat cycle which consists of two isobaric paths and two isochoric paths, as given in Fig. \ref{PV-dia}.
Note that, from Eqs. (\ref{bhentr}) and (\ref{vol}), we can infer that the entropy $S$ is a function
of the thermodynamic volume $V$. This suggests $dS\sim dV$ and thus the isochoric or adiabatic paths are the same. It is easy to calculate the total work done in this heat cycle as
\begin{eqnarray}
W=\oint PdV&=&(P_1-P_4)(V_2-V_1),\nonumber\\
 &=&\frac{4\pi}{3}(P_1-P_4)\left(r^3_{+2}e^{\frac{k}{2r_{+2}}}-r^3_{+1}e^{\frac{k}{2r_{+1}}}\right).
\end{eqnarray}
Whereas, the amount of the input heat is calculated as
\begin{eqnarray}
Q_H&=&\int^{T_2}_{T_1}C_P(P_1,T)dT=\int^{r_{+2}}_{r_{+1}}\frac{\partial M}{\partial r_+}dr_+,\nonumber\\
&=&\frac{4\pi}{3}P_1r_+e^{\frac{k}{2r_+}}\left[r^2_++\frac{3}{8\pi P_1}+\frac{3m^2}{8\pi P_1}\left(\frac{cc_1r_+}{2}+c^2c_2\right)\right]\Big|^{r_{+2}}_{r_{+1}}.
\end{eqnarray}
The heat engine efficiency of the black hole thus is given by
\begin{eqnarray}
\eta&=&\frac{W}{Q_H}=\left(1-\frac{P_4}{P_1}\right)\frac{r^3_{+2}e^{\frac{k}{2r_{+2}}}-r^3_{+1}e^{\frac{k}{2r_{+1}}}}{r_+e^{\frac{k}{2r_+}}\left[r^2_++\frac{3}{8\pi P_1}+\frac{3m^2}{8\pi P_1}\left(\frac{cc_1r_+}{2}+c^2c_2\right)\right]\Big|^{r_{+2}}_{r_{+1}}}.\label{heat-eff}
\end{eqnarray}
Here, by solving Eq. (\ref{bhentr}) we can express  $r_{+2}$ and $r_{+1}$ as the functions of the corresponding entropies as
\begin{equation}
r_{+2}=r_{+2}(S_2),\ \ r_{+1}=r_{+1}(S_1).
\end{equation}
Or, by solving Eq. (\ref{bhtemp}) we can express them as the functions of the corresponding temperature and pressure as
\begin{equation}
r_{+2}=r_{+2}(T_H,P_1),\ \ r_{+1}=r_{+1}(T_C,P_4),
\end{equation}
where we have used $T_C\equiv T_4$ and $T_H\equiv T_2$. In general, because the expressions for the black hole entropy and the black hole temperature are quite complex, it is very difficult to get explicitly these functions. However, we can do this in some limits. For the case $k/r_+\ll 1$, we have the following approximation for the functions $r_{+2}=r_{+2}(S_2)$ and $r_{+1}=r_{+1}(S_1)$ as
\begin{equation}
r_{+2}=\frac{k}{2}\left(\sqrt{1+\frac{4S_2}{\pi k^2}}-1\right),\ \ \ \ r_{+1}=\frac{k}{2}\left(\sqrt{1+\frac{4S_1}{\pi k^2}}-1\right).
\end{equation}
Let us consider the heat cycle in the limit of the high temperature and high pressure in which the event horizon
radius $r_+$ of the black hole can be expanded as
\begin{equation}
r_+=\sum^{\infty}_{n=0}\epsilon^nr_n,\ \ \epsilon=\frac{1}{8\pi
P}.
\end{equation}
Using this expansion, one can solve perturbatively for $r_+$ from Eq. (\ref{bhtemp}). At the lowest-order approximation $(n=0)$, we have
\begin{equation}
r_+=r_0=\frac{T}{2P}+\frac{k}{6}.\label{loapp-r}
\end{equation}
At this approximation, the output work and the net inflow of the black hole heat engine are given by
\begin{eqnarray}
W&=&\frac{4\pi}{3}(P_1-P_4)\left(r^3_{02}e^{\frac{k}{2r_{02}}}-r^3_{01}e^{\frac{k}{2r_{01}}}\right)\equiv W_0,\nonumber\\
Q_H&=&\frac{4\pi}{3}P_1\left(r^3_{02}e^{\frac{k}{2r_{02}}}-r^3_{01}e^{\frac{k}{2r_{01}}}\right)\equiv Q_{H0},
\end{eqnarray}
where
\begin{equation}
r_{01}=\frac{T_C}{2P_4}+\frac{k}{6},\ \ \ \ r_{02}=\frac{T_H}{2P_1}+\frac{k}{6}.
\end{equation}
As a result, at the lowest-order approximation the heat engine efficiency of the black hole becomes
\begin{eqnarray}
\eta=\frac{W}{Q_H}=1-\frac{P_4}{P_1}.
\end{eqnarray}
At the leading-order approximation ($n=1$), we have
\begin{equation}
r_+=r_0+\epsilon r_1=r_0+\frac{k-2r_0}{16\pi P_1r^2_0}\left[1+m^2\left(\frac{cc_1r_0}{2}+c^2c_2\right)\right]-\frac{m^2cc_1}{16\pi P_1}.
\end{equation}
We can easily obtain the output work and the net inflow of the heat at this approximation as
\begin{eqnarray}
W&=&W_0+\frac{W_1}{8\pi P_1},\nonumber\\
Q_H&=&Q_{H0}+\frac{Q_{H1}}{8\pi P_1},
\end{eqnarray}
where
\begin{eqnarray}
W_1&=&\frac{2\pi(P_1-P_4)}{3}\left[r_{02}r_{12}(6r_{02}-k)e^{\frac{k}{2r_{02}}}-r_{01}r_{11}(6r_{01}-k)e^{\frac{k}{2r_{01}}}\right],\nonumber\\
Q_{H1}&=&\frac{P_1}{P_1-P_4}W_1+4\pi P_1\left\{r_{02}e^{\frac{k}{2r_{02}}}\left[1+m^2\left(\frac{cc_1r_{02}}{2}+c^2c_2\right)\right]-r_{01}e^{\frac{k}{2r_{01}}}\left[1+m^2\left(\frac{cc_1r_{01}}{2}+c^2c_2\right)\right]\right\},\nonumber\\
\end{eqnarray}
with
\begin{eqnarray}
r_{12}&=&\frac{k-2r_{02}}{2r^2_{02}}\left[1+m^2\left(\frac{cc_1r_{02}}{2}+c^2c_2\right)\right]-\frac{m^2cc_1}{2},\nonumber\\
r_{11}&=&\frac{k-2r_{01}}{2r^2_{01}}\left[1+m^2\left(\frac{cc_1r_{01}}{2}+c^2c_2\right)\right]-\frac{m^2cc_1}{2}.
\end{eqnarray}
Then, the heat engine efficiency of the black hole corresponding to the leading-order approximation is given by
\begin{eqnarray}
\eta&=&\left(1-\frac{P_4}{P_1}\right)\left[1-\frac{1}{8\pi P_1}\left(\frac{Q_{H1}}{Q_0}-\frac{W_1}{W_0}\right)\right]+\mathcal{O}\left(\frac{1}{P^2_1}\right).
\end{eqnarray}

\begin{figure}[t]
% Requires \usepackage{graphicx}\
 \centering
\begin{tabular}{cc}
\includegraphics[width=0.45 \textwidth]{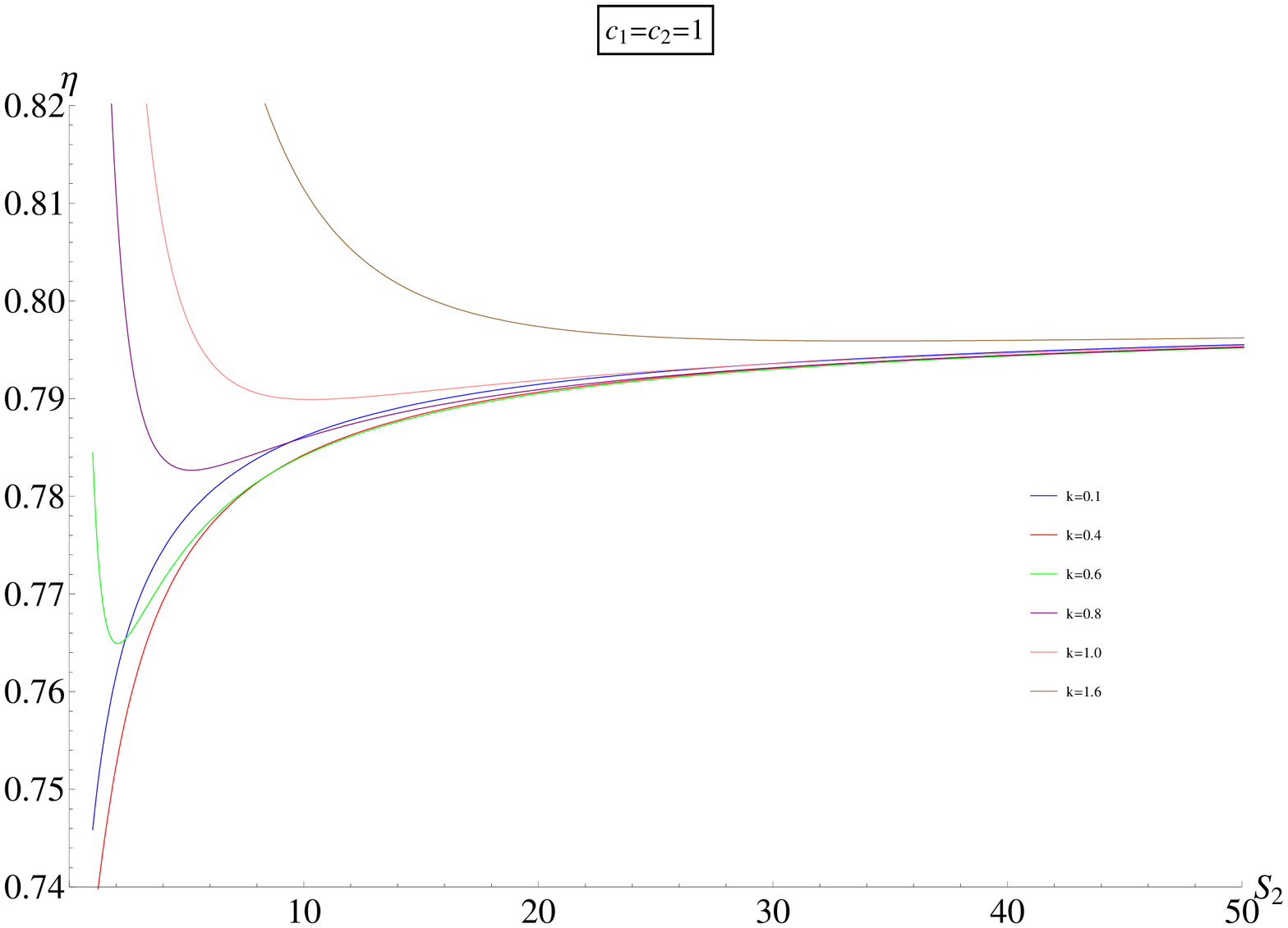}
\hspace*{0.05\textwidth}
\includegraphics[width=0.45 \textwidth]{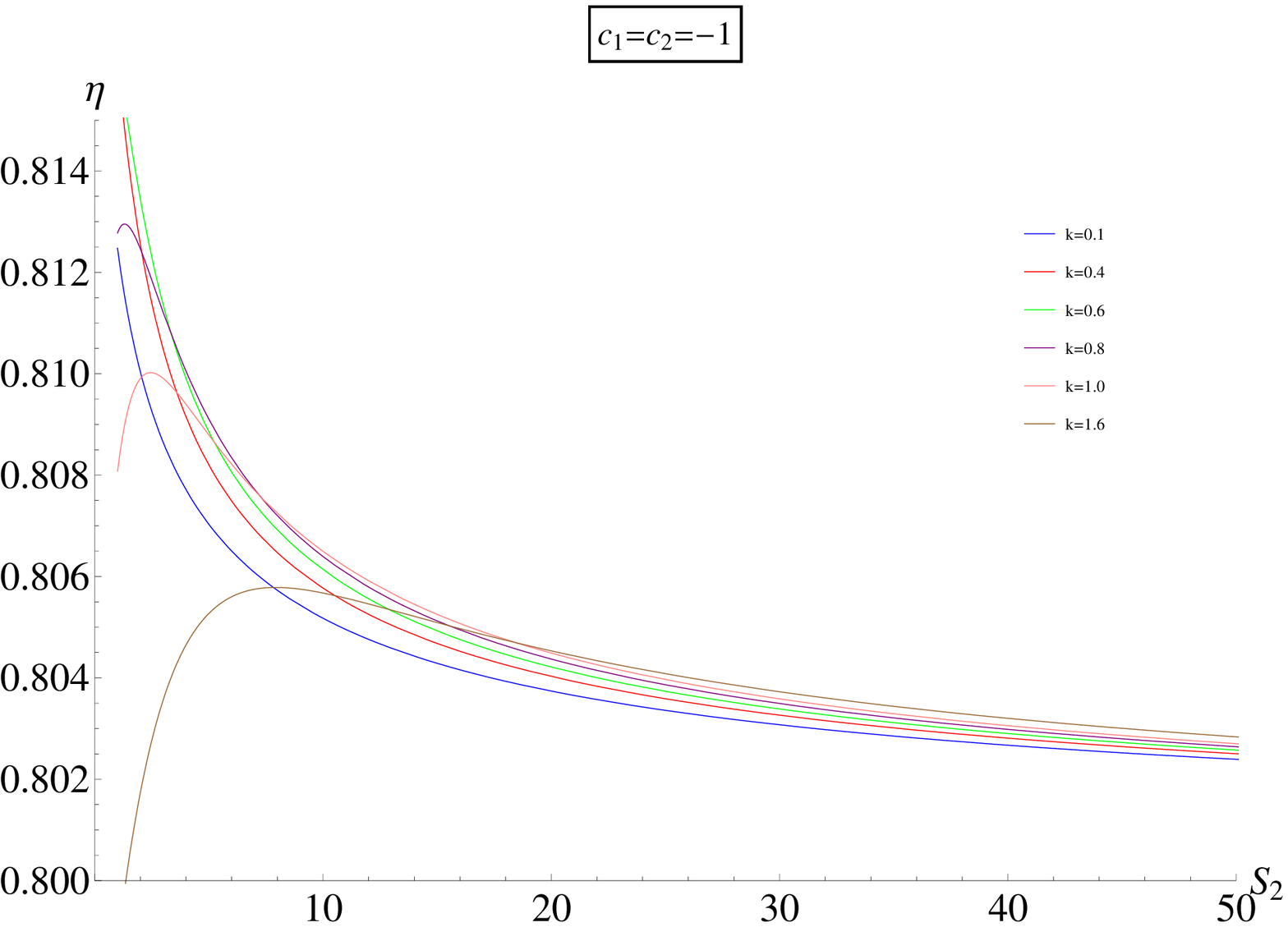}\\
\includegraphics[width=0.45 \textwidth]{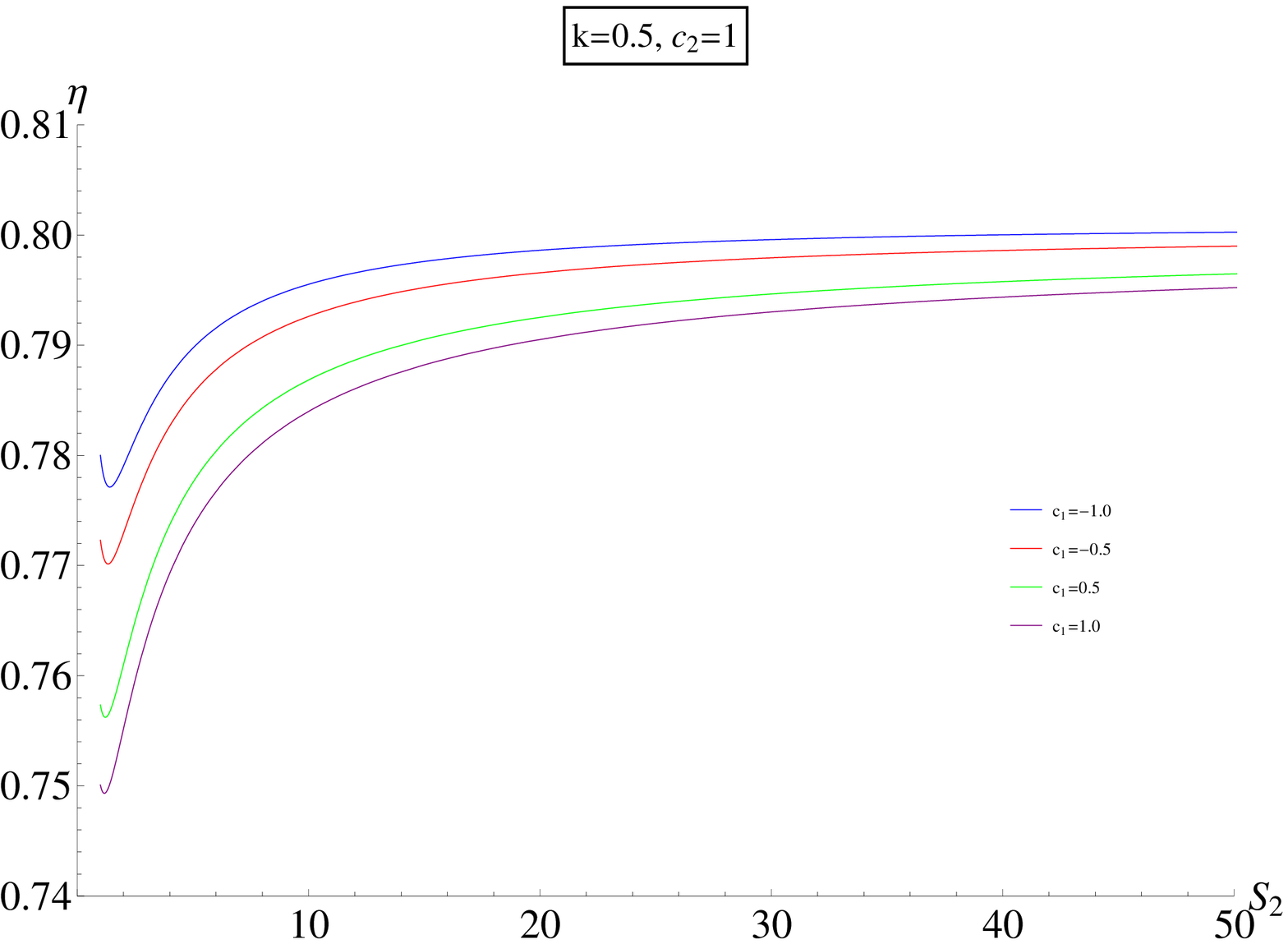}
\hspace*{0.05\textwidth}
\includegraphics[width=0.45 \textwidth]{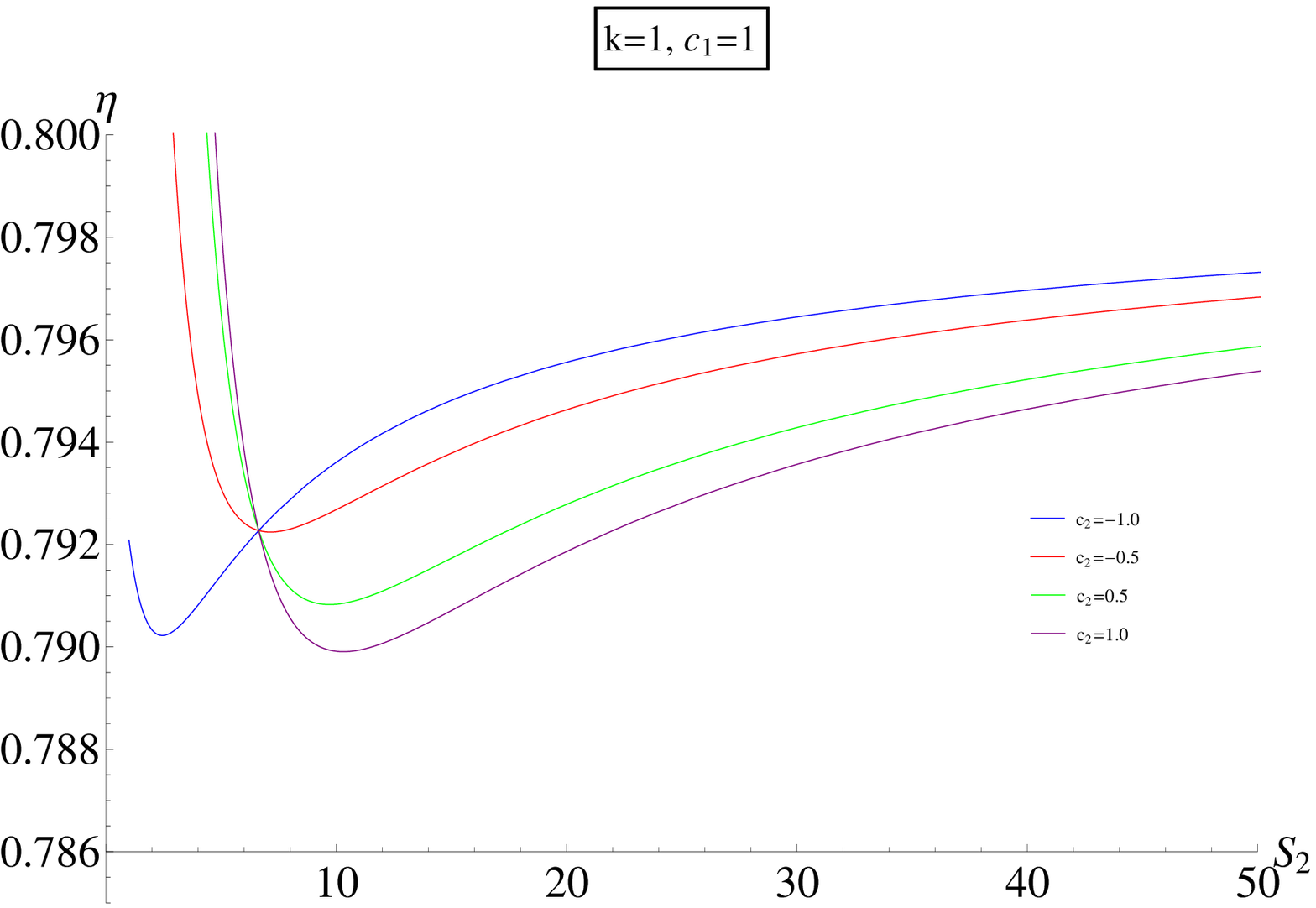}
\end{tabular}
  \caption{Heat engine efficiency $\eta$ of the black hole as a function of the entropy $S_2$, at $S_1=P_4=m=c=1$ and $P_1=5$.}\label{eta-S2}
\end{figure}

\begin{figure}[t]
% Requires \usepackage{graphicx}\
 \centering
\begin{tabular}{cc}
\includegraphics[width=0.5 \textwidth]{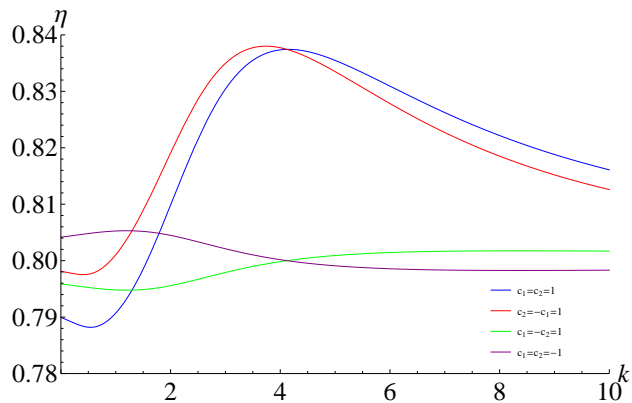}
\end{tabular}
  \caption{Heat engine efficiency as a function of the non-linear parameter $k$, at $S_1=P_4=m=c=1$, $S_2=15$ and $P_1=5$.}\label{eta-k}
\end{figure}

Now we analyze the behavior of the heat engine efficiency $\eta$ as a function of the entropy $S_2$ and the pressure $P_1$, corresponding to the heat cycle given in Fig. \ref{PV-dia}, for the various values of the non-linear and massive gravity parameters. By employing Eqs. (\ref{bhentr}) and (\ref{heat-eff}), one can generate the parametric plots of the entropy $S_2$ and the heat engine efficiency $\eta$ , with the entropy $S_1$ and the pressures $P_{1,2}$ kept fixed all, which are given in Fig. \ref{eta-S2}.
From this figure, we observe that the behavior of the heat engine efficiency is crucially dependent on the non-linear and massive gravity parameters. With the proper parameters, the heat engine efficiency is a monotonously increasing/decreasing function with the growth of the entropy $S_2$. This suggests that the bigger black holes have the larger/smaller efficiency of the heat engine. Whereas, with other proper combinations of the parameters, they lead to the heat engine efficiency curve which has a global maximum/minimum value. This means that there exits a finite value of the entropy $S_2$ at which the heat engine of the black hole works at the highest or lowest efficiency. In the limit of that the entropy $S_2$ goes to the infinity, the heat engine efficiency should approach $\eta=1-P_4/P_1$. 
\begin{figure}[t]
% Requires \usepackage{graphicx}\
 \centering
\begin{tabular}{cc}
\includegraphics[width=0.45 \textwidth]{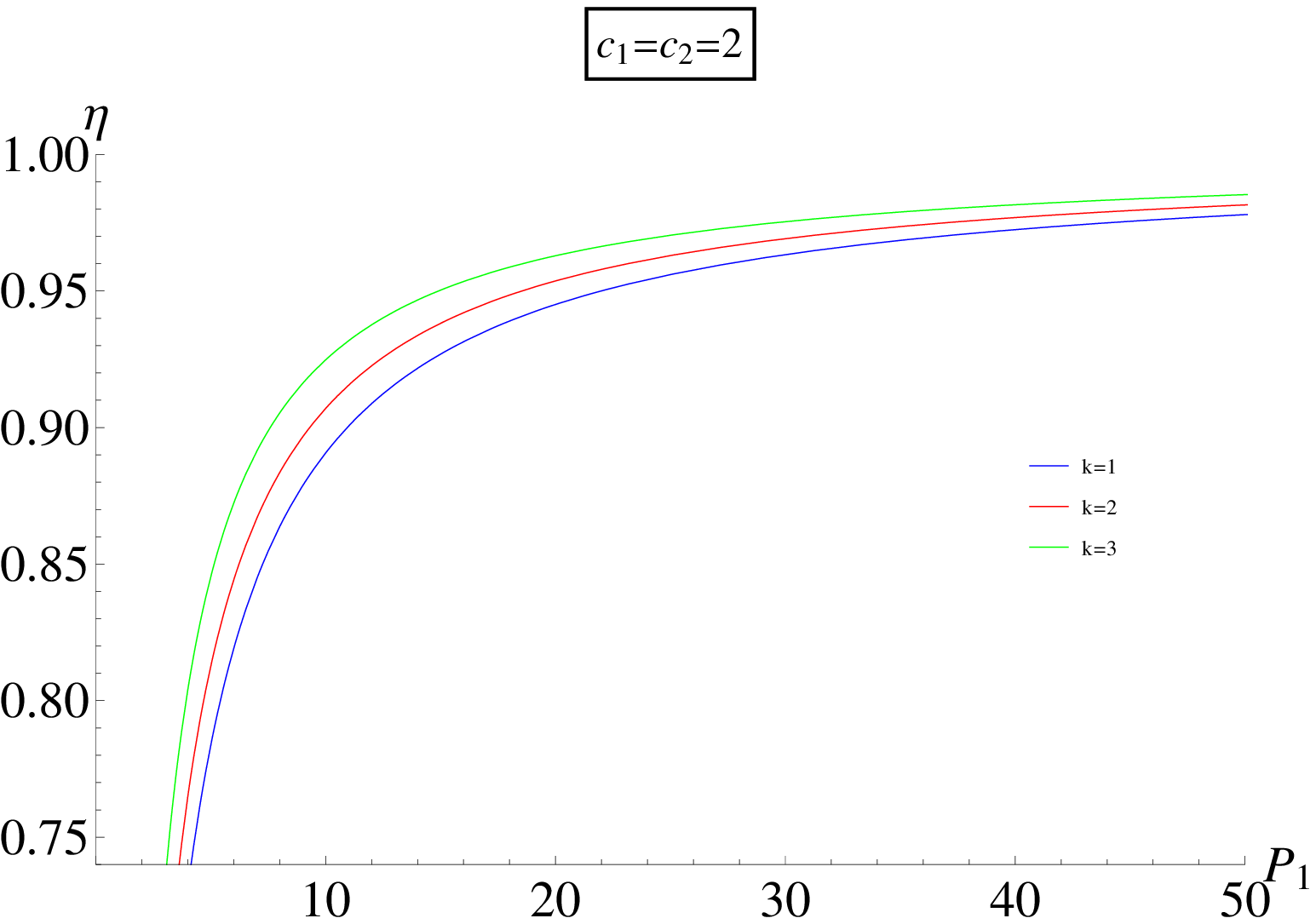}
\hspace*{0.05\textwidth}
\includegraphics[width=0.45 \textwidth]{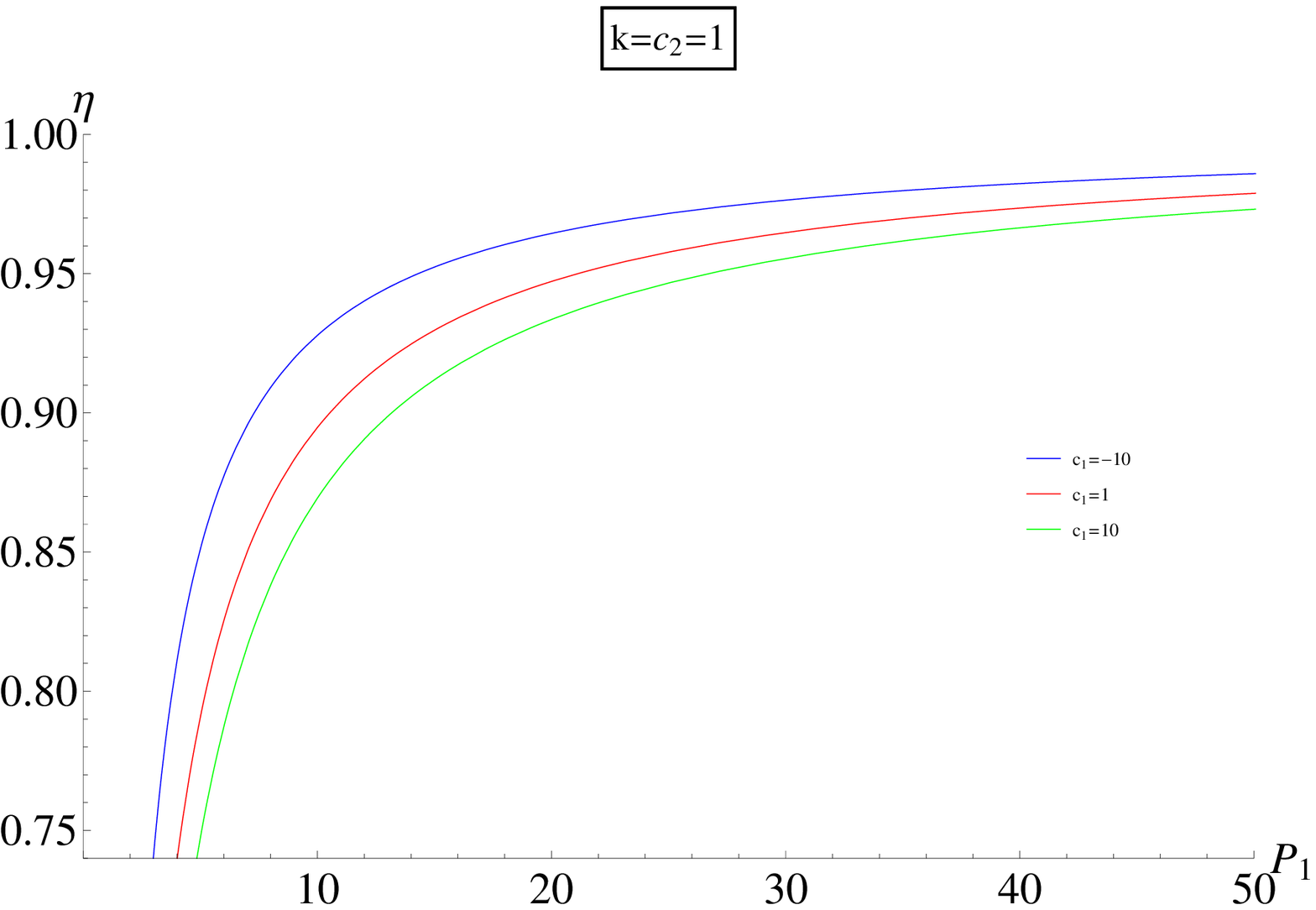}
\end{tabular}
  \caption{Heat engine efficiency as a function of the pressure $P_1$, at $S_1=P_4=m=c=1$ and $S_1=15$.}\label{eta-P1}
\end{figure}
In addition, the heat engine efficiency is plotted against the pressure $P_1$, with the entropies $S_{1,2}$ and the pressure $P_4$ kept fixed all, given in Fig. \ref{eta-P1}. From this figure, we found that
increasing the pressure makes the increasing of the heat engine efficiency. In the limit of that the pressure $P_1$ goes to the infinity, the efficiency should approach $\eta=1$. From the mentioned two figures, we can see that the heat engine efficiency almost decreases as the massive gravity parameters increase. In order to see more explicitly the effect of the non-linear parameter $k$ on the heat engine efficiency $\eta$, we plot $\eta$ as a function of $k$ in Fig. \ref{eta-k}.
We observe that depending the value region of the non-linear parameter $k$ as well as the sign of the massive gravity parameters, increasing $k$ leads to the increasing or decreasing of the heat engine efficiency.
\begin{figure}[t]
% Requires \usepackage{graphicx}\
 \centering
\begin{tabular}{cc}
\includegraphics[width=0.45 \textwidth]{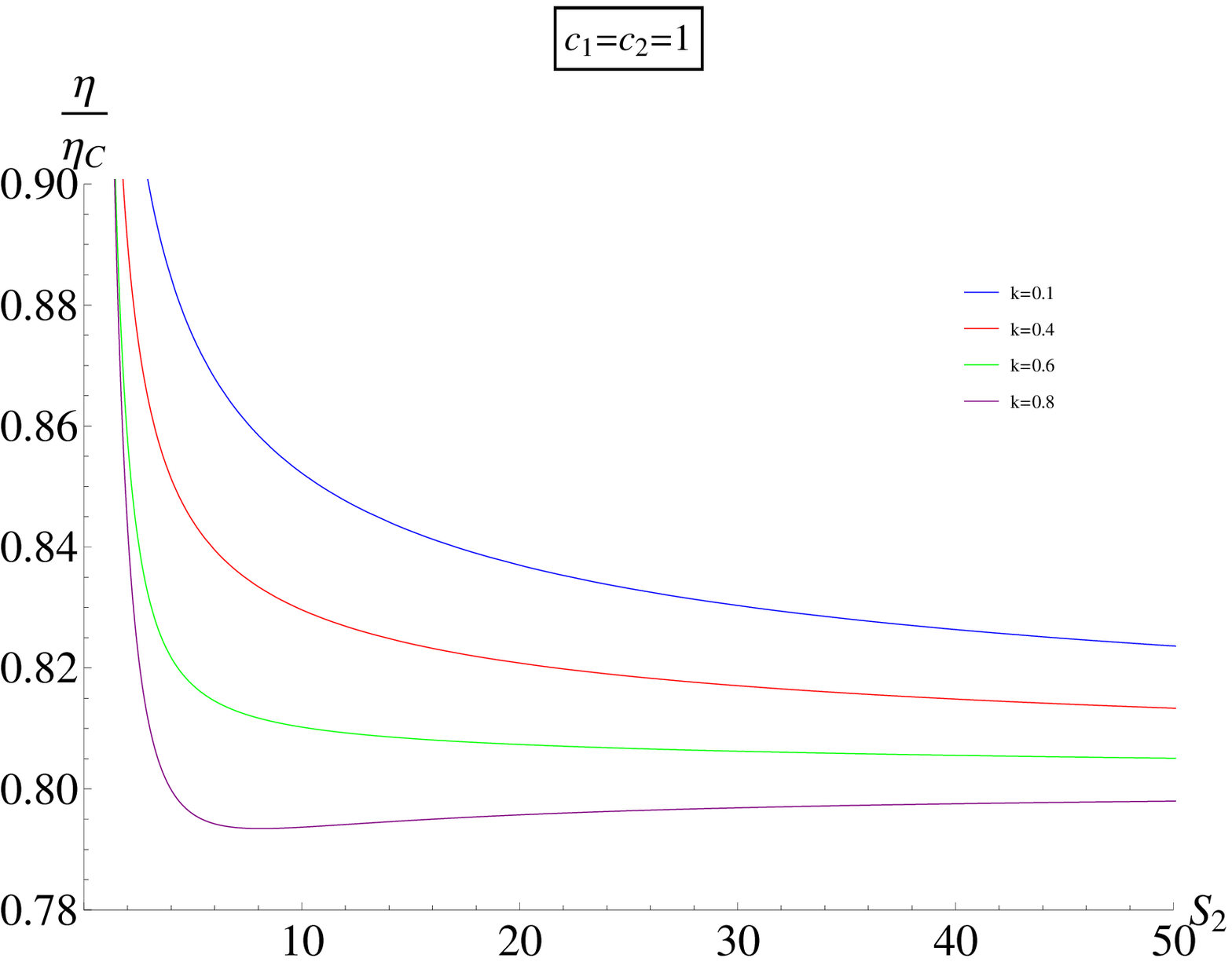}
\hspace*{0.05\textwidth}
\includegraphics[width=0.45 \textwidth]{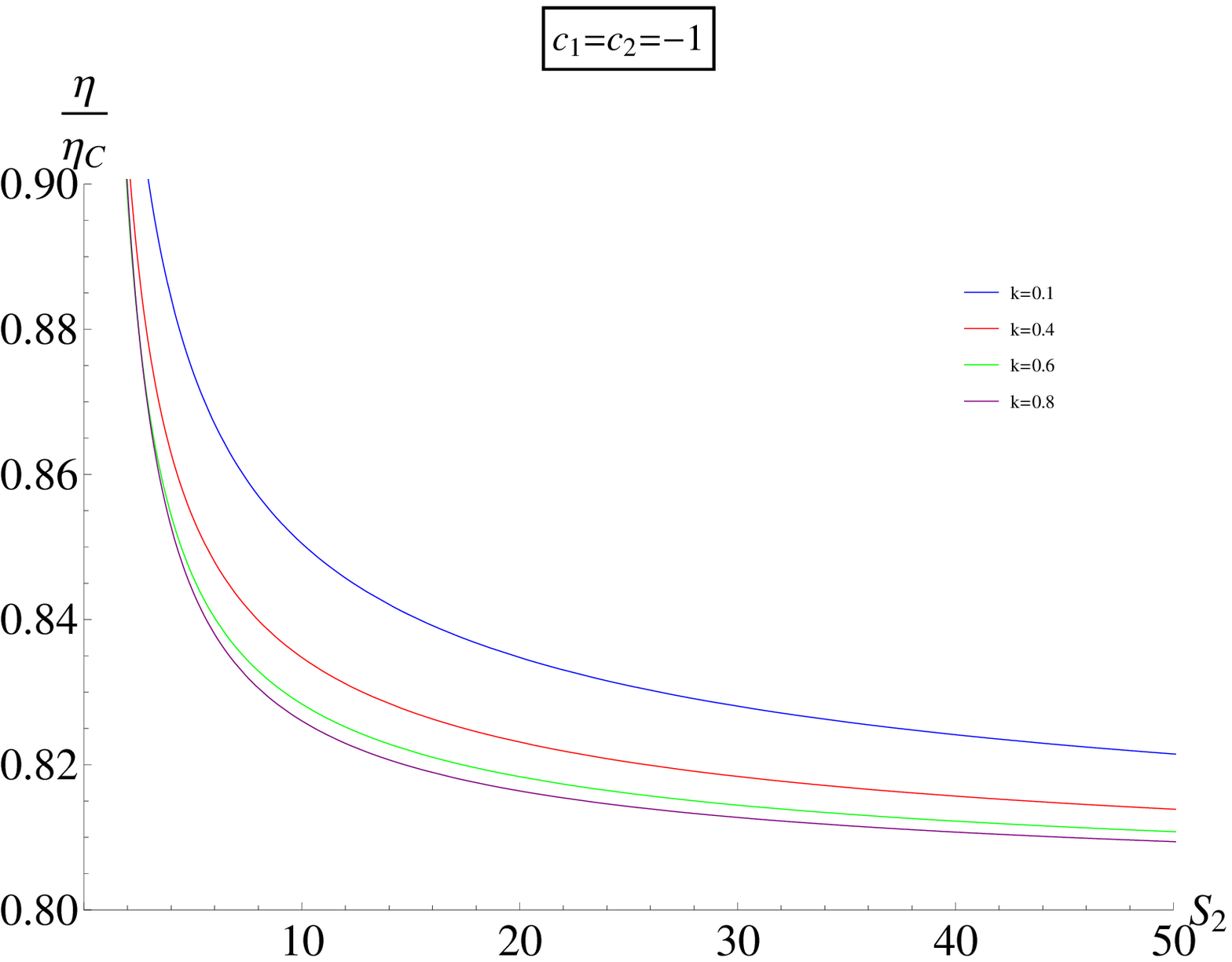}\\
\includegraphics[width=0.45 \textwidth]{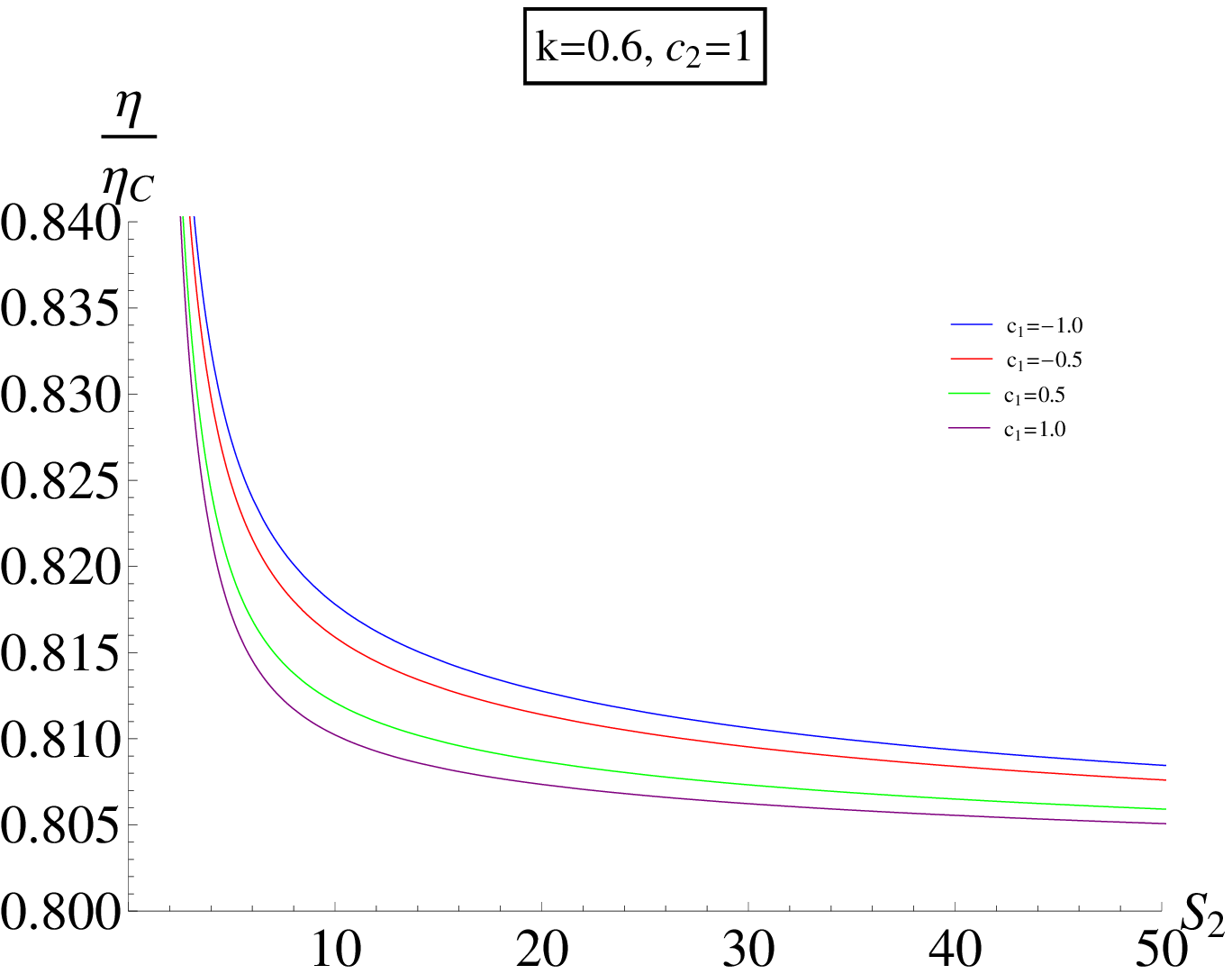}
\hspace*{0.05\textwidth}
\includegraphics[width=0.45 \textwidth]{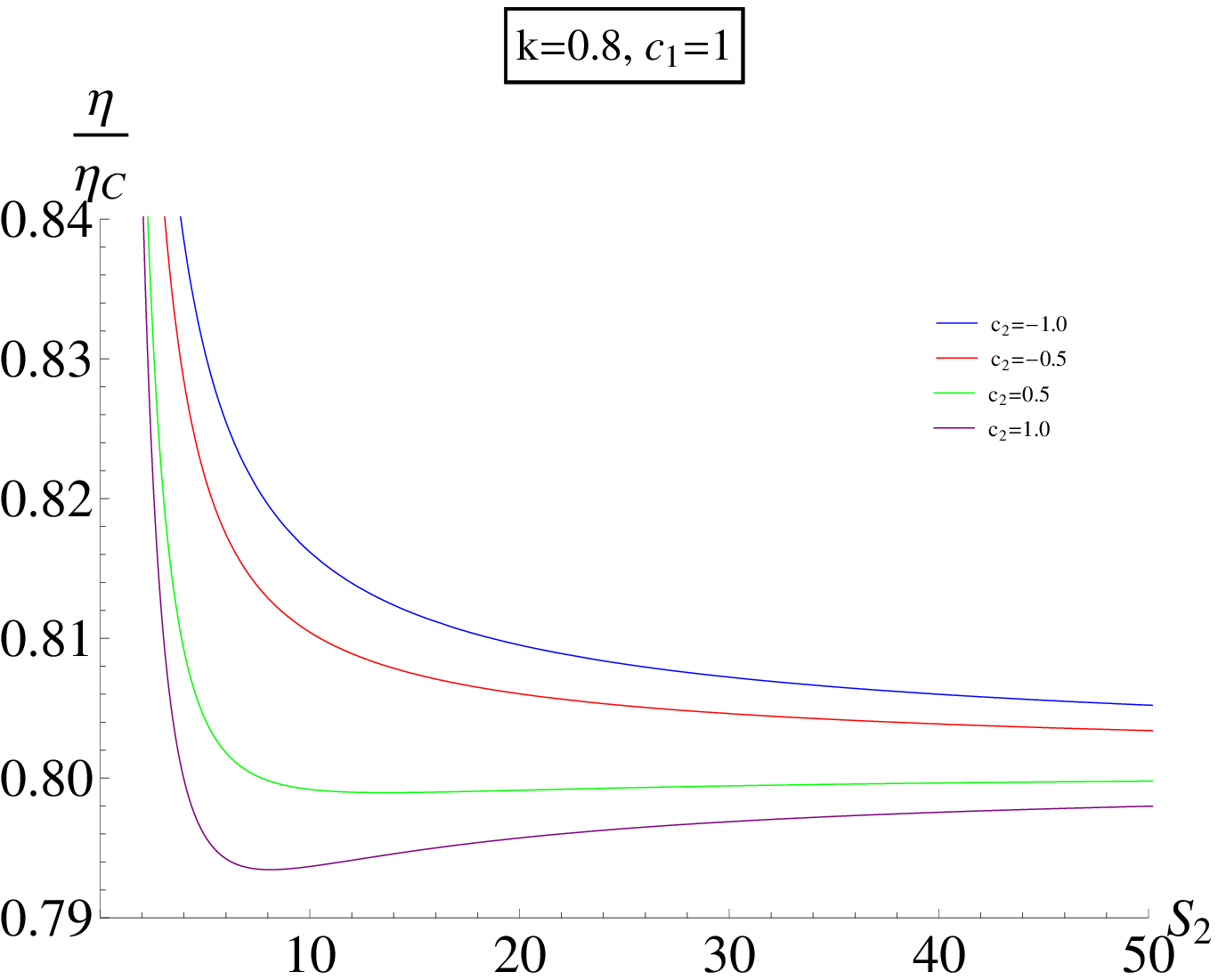}
\end{tabular}
\caption{The ratio $\eta/\eta_C$ as a function of the entropy $S_2$, at $S_1=P_4=m=c=1$, and $P_1=5$.}\label{comp-S2}
\end{figure}
\begin{figure}[t]
% Requires \usepackage{graphicx}\
 \centering
\begin{tabular}{cc}
\includegraphics[width=0.45 \textwidth]{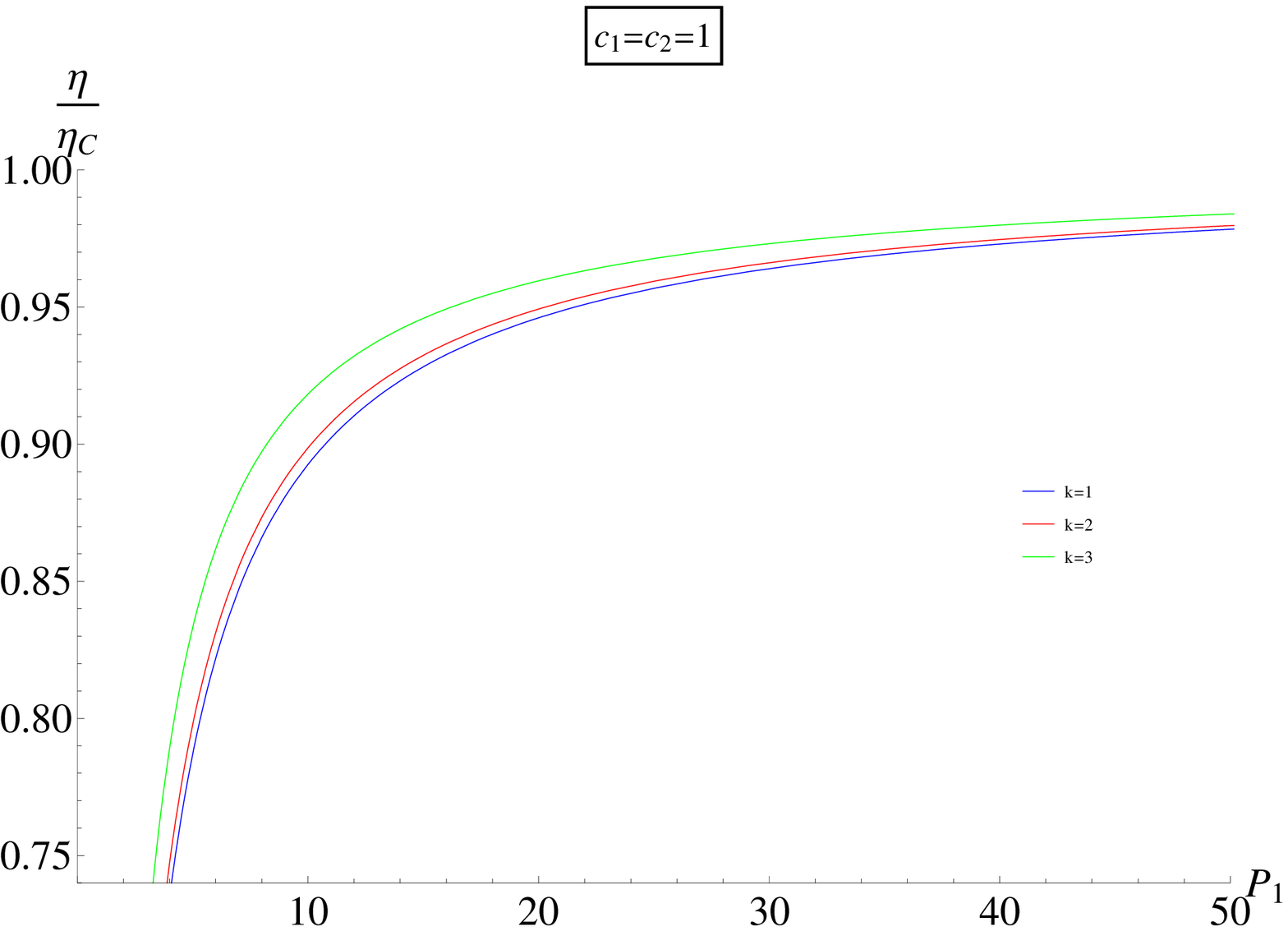}
\hspace*{0.05\textwidth}
\includegraphics[width=0.45 \textwidth]{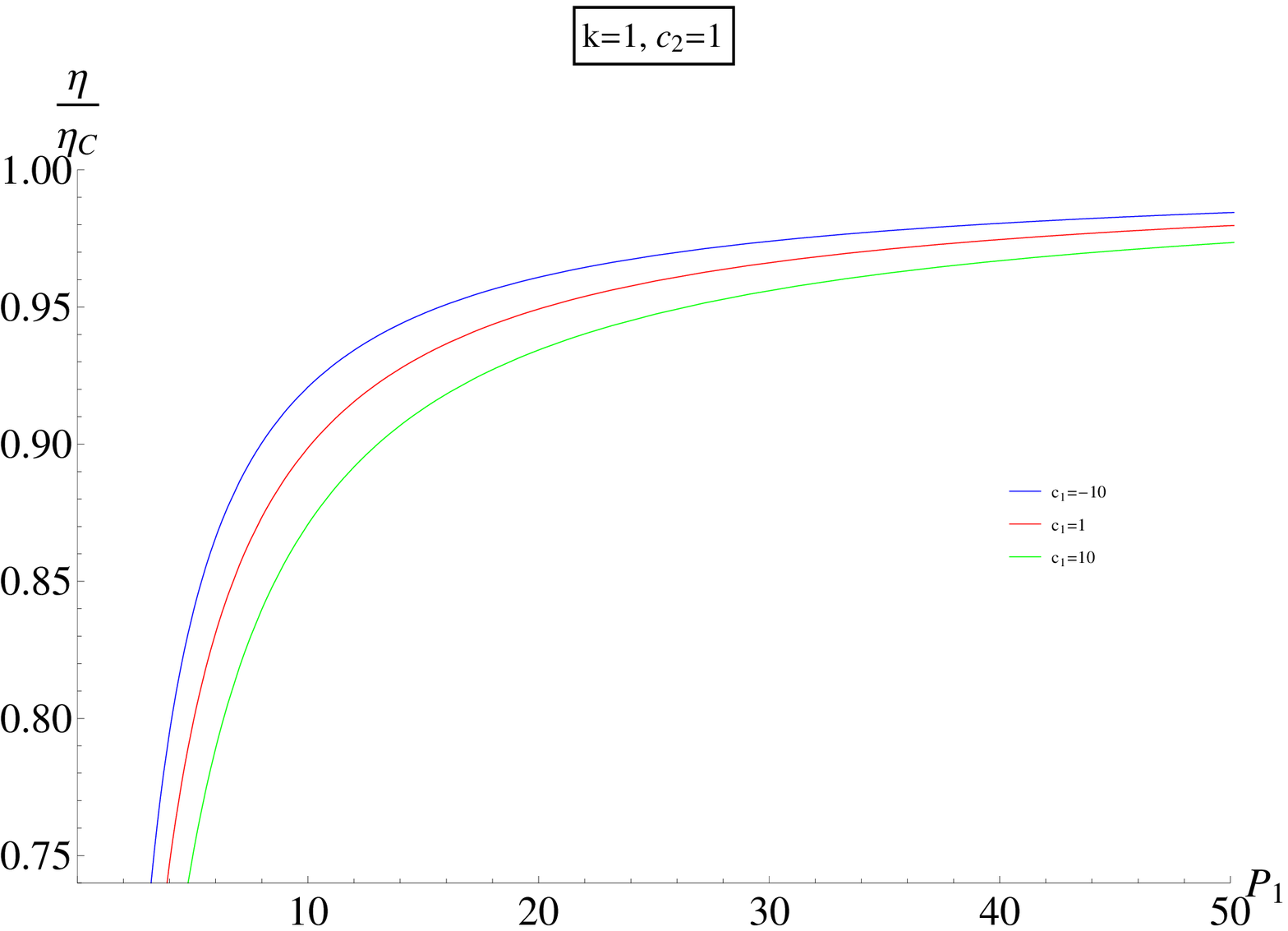}
\end{tabular}
\caption{The ratio $\eta/\eta_C$ as a function of the pressure $P_1$, at $S_1=P_4=m=c=1$, and $S_2=15$.}\label{comp-P1}
\end{figure}

\begin{figure}[t]
% Requires \usepackage{graphicx}\
 \centering
\begin{tabular}{cc}
\includegraphics[width=0.5 \textwidth]{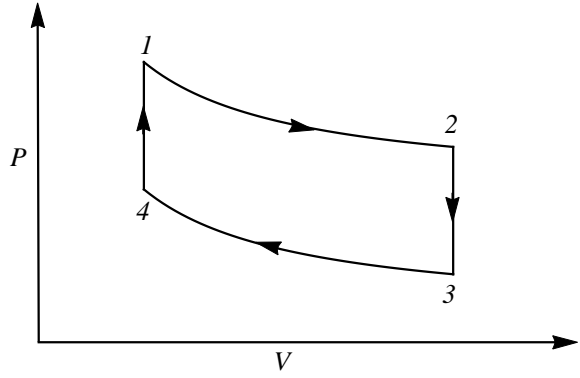}
\end{tabular}
  \caption{A Carnot heat cycle with two isothermal paths and two isochoric paths.}\label{C-heat-cycle}
\end{figure}

Let us compare the heat engine efficiency $\eta$ with the Carnot efficiency $\eta_C$ which is the maximum value. The black hole as a heat engine with the Carnot efficiency is described by the heat cycle including a pair of isothermal paths connected to each other by a pair of either isochoric paths, as shown in Fig. \ref{C-heat-cycle}.
It is easily to calculate the amount of the input heat $Q_H$ and the amount of the exhaust heat $Q_C$ as
\begin{eqnarray}
Q_H&=&T_H(S_2-S_1),\nonumber\\
Q_C&=&T_C(S_3-S_4).
\end{eqnarray}
From Eqs. (\ref{bhentr}) and (\ref{vol}), we can find $S_2-S_1=S_3-S_4$ for the heat cycle given in \ref{C-heat-cycle}. As a result, the Carnot efficiency $\eta_C$ is given by
\begin{equation}
\eta_C=1-\frac{Q_C}{Q_H}=1-\frac{T_C}{T_H}.
\end{equation}
In Figs. \ref{comp-S2} $\&$ \ref{comp-P1}, we plot the ratio $\eta/\eta_C$ as a function of the entropy $S_2$ and the pressure $P_1$. We find that the heat engine efficiency $\eta$ is close the Carnot efficiency $\eta_C$ in the region of the entropy $S_2$ near $S_1$. The ratio $\eta/\eta_c$ is a monotonously increasing function with the growth of the pressure $P_1$. This implies that the heat engine efficiency should approach the Carnot efficiency as the pressure $P_1$ goes to the infinity. Furthermore, the change of the non-linear and massive gravity parameters affect significantly on the ratio $\eta/\eta_C$. More specifically, increasing the non-linear parameter $k$ makes either increasing or decreasing the ratio $\eta/\eta_C$, which are dependent on the value region of $k$. Whereas, increasing the massive gravity parameters leads to a lower ratio $\eta/\eta_C$. 

\section{\label{JT-expansion} Joule-Thomson expansion of non-linear
charged AdS black hole}

\begin{figure}[t]
% Requires \usepackage{graphicx}\
 \centering
\begin{tabular}{cc}
\includegraphics[width=0.45 \textwidth]{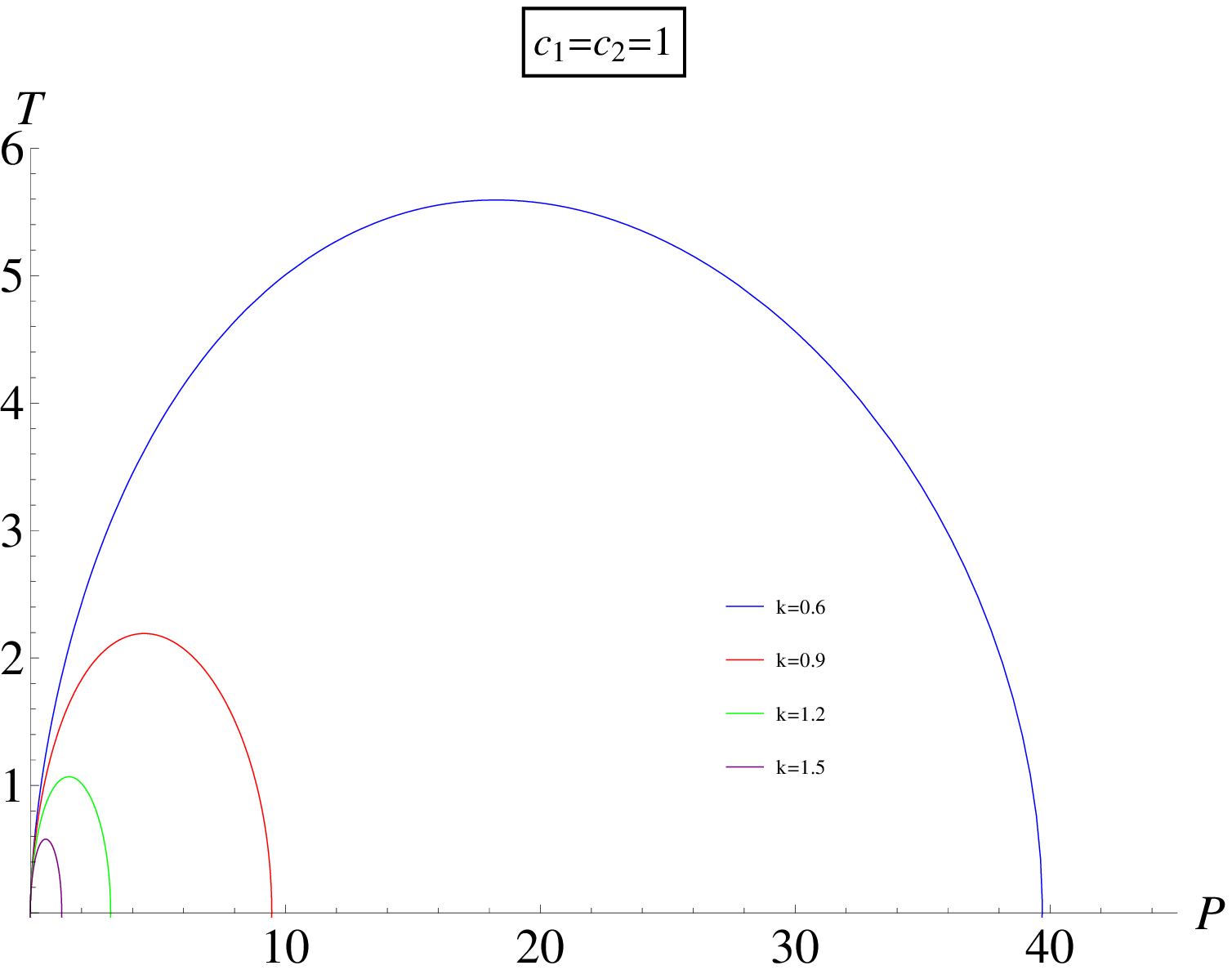}
\hspace*{0.05\textwidth}
\includegraphics[width=0.45 \textwidth]{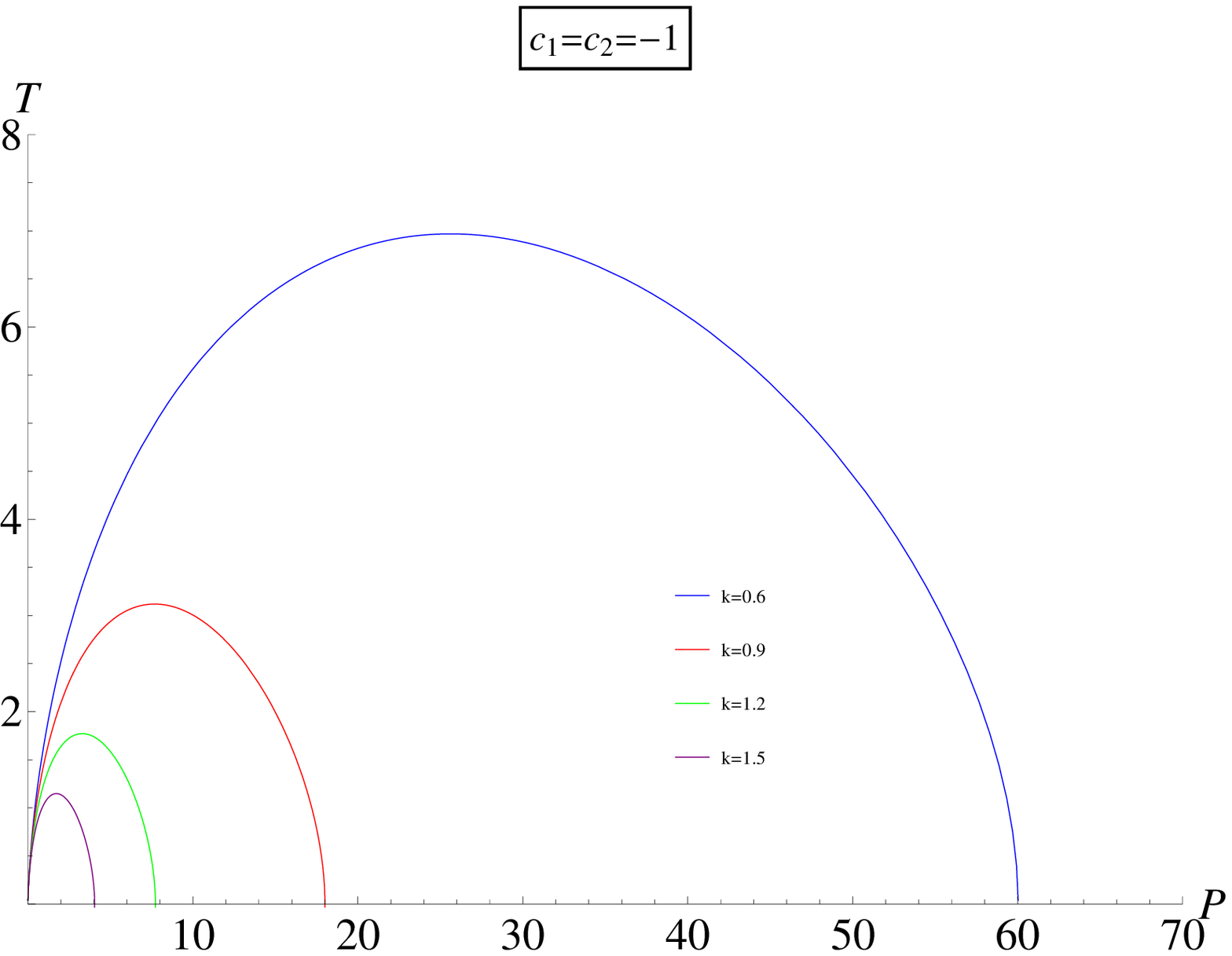}\\
\includegraphics[width=0.45 \textwidth]{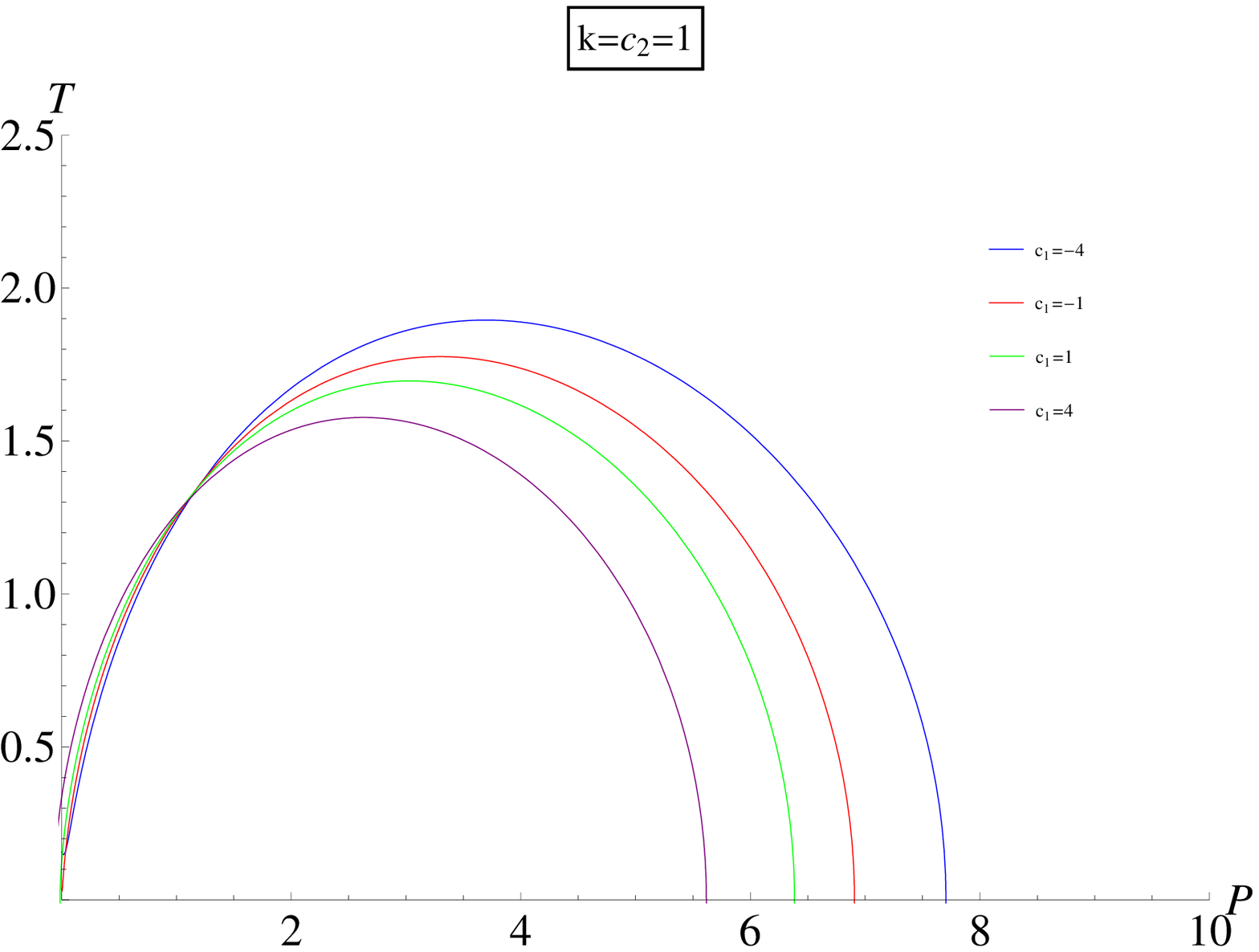}
\hspace*{0.05\textwidth}
\includegraphics[width=0.45 \textwidth]{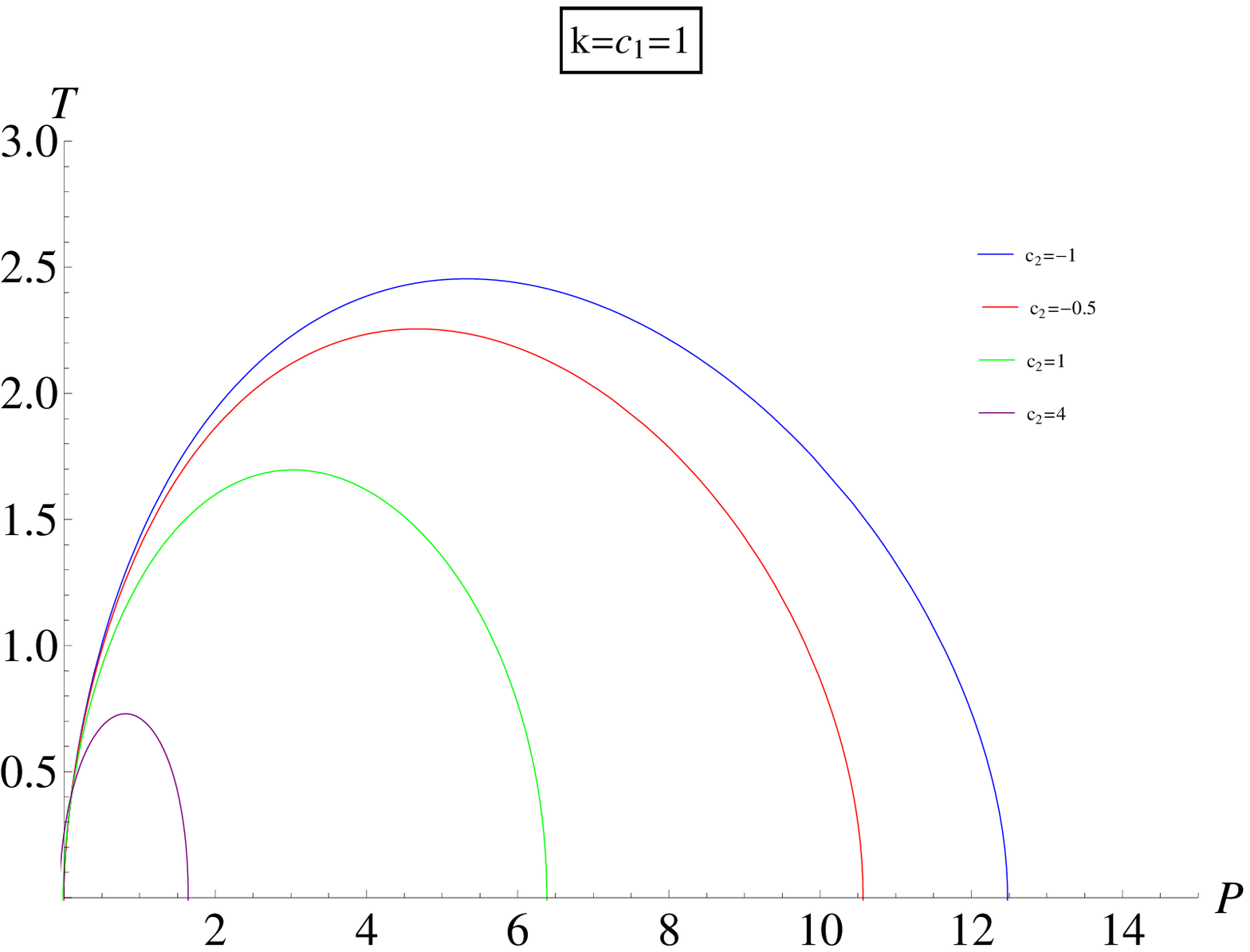}
\end{tabular}
\caption{The constant mass curves are plotted for various values of the non-linear parameter $k$ and the massive gravity coupling parameters $c_{1,2}$, at $m=c=1$, and $M=5$.}\label{P-T-fig}
\end{figure}

In this section, we will study the isenthalpy process or the Joule-Thomson expansion of the non-linear
charged AdS black hole in the presence of graviton mass. The Joule-Thomson expansion of the black hole is described by the constant mass curves in the $T-P$ diagram. From Eqs. (\ref{enthalpy}) and (\ref{bhtemp}), we can express the pressure and the temperature as the functions of the black hole mass $M$ and its event horizon radius $r_+$ as 
\begin{eqnarray}
P(M,r_+)&=&\frac{3}{16\pi r^3_+}\left[4Me^{-\frac{k}{2r_+}}-2(1+m^2c^2c_2)r_+-m^2cc_1r^2_+\right],\label{P-eq} \\
T(M,r_+)&=&\frac{1}{8\pi r^3_+}\left[2M(6r_+-k)e^{-\frac{k}{2r_+}}-4(1+m^2c^2c_2)r^2_+-m^2cc_1r^3_+\right].\label{T-eq}
\end{eqnarray}
In principle, in order to find  the function $T=T(M,P)$, one needs to solve Eq. (\ref{P-eq}) to obtain the event horizon radius as a function of the pressure $P$ and the black hole mass $M$ then substituting it into Eq. (\ref{T-eq}). Because of the complexity of Eq. (\ref{P-eq}), it is in general a difficult task. However, employing these equations can generate the parametric plots of the pressure $P$ and the temperature $T$. In Fig. \ref{P-T-fig}, we show the constant mass curves for various values of the non-linear parameter $k$ and the massive gravity coupling parameters $c_{1,2}$.
From this figure, we observe that with the same mass, as the non-linear parameter $k$ and the massive gravity coupling parameters $c_{1,2}$ increase, the constant mass curves tend to shrink towards the lower pressure and temperature.

The pressure always decreases in the Joule-Thomson expansion. But, the temperature can increase or decrease. The increasing of the temperature in the Joule-Thomson expansion corresponds to the appearance of the heating process, associated with the the negative slope of the constant mass curves. Whereas, the decreasing of the temperature corresponds to the appearance of the cooling process, associated with the the positive slope of the constant mass curves. An essential physical quantity, which characterizes the Joule-Thomson expansion, is the change of the temperature with respect to the pressure along the constant mass curves, called the Joule-Thomson coefficient. This quantity not only describes that the Joule-Thomson expansion is fast or slow, but also it allows us to determine whether the heating or the cooling process will appear in this expansion. The Joule-Thomson coefficient $\mu$ is thus defined as
\begin{eqnarray}
\mu&=&\left(\frac{\partial T}{\partial P}\right)_M=\left(\frac{\partial T}{\partial r_+}\right)_M\Big/\left(\frac{\partial P}{\partial r_+}\right)_M,\nonumber\\
&=&\frac{8r^2_+\left[e^{\frac{k}{2r_+}}(1+m^2c^2c_2)r_+-6M\right]+24kMr_+-2k^2M}{6M(k-6r_+)+3e^{\frac{k}{2r_+}}\left[4(1+m^2c^2c_2)+m^2cc_1r_+\right]r^2_+}.\label{JT-coeff}
\end{eqnarray}
It is interesting that we can express the Joule-Thomson coefficient $\mu$ in terms of the temperature, the thermodynamic volume and the heat capacity. For the Joule-Thomson expansion, we have $dM=0$, which leads to the following relation
\begin{equation}
T\left(\frac{\partial S}{\partial P}\right)_M+V=0.\label{dM}
\end{equation}
In addition, from Eqs. (\ref{bhtemp}) and (\ref{bhentr}), the black hole entropy can be understood as a state function, $S=S(T,P)$, and thus we find 
\begin{equation}
\left(\frac{\partial S}{\partial P}\right)_M=\left(\frac{\partial
S}{\partial P}\right)_T+\left(\frac{\partial S}{\partial
T}\right)_P\left(\frac{\partial T}{\partial P}\right)_M.\label{dS}
\end{equation}
With these two relation, one can easily obtain
\begin{equation}
T\left[\left(\frac{\partial S}{\partial
P}\right)_T+\left(\frac{\partial S}{\partial
T}\right)_P\left(\frac{\partial T}{\partial
P}\right)_M\right]+V=0.
\end{equation}
By using the Maxwell relation $\left(\partial S/\partial
P\right)_T=-\left(\partial V/\partial T\right)_P$ and the definition of the heat capacity $C_P=T\left(\partial S/\partial T\right)_P$, we derive 
\begin{equation}
\mu=\left(\frac{\partial T}{\partial
P}\right)_M=\frac{1}{C_P}\left[T\left(\frac{\partial V}{\partial
T}\right)_P-V\right].\label{JT-coeff-2}
\end{equation}
The expressions for the Joule-Thomson coefficient $\mu$ are given at Eqs. (\ref{JT-coeff}) and (\ref{JT-coeff-2}) are equivalent.

The inversion points $(P_i,T_i)$ form an inversion curve that divides the $T-P$ graph into the cooling and heating regions. The inversion points $(P_i,T_i)$ are obtained by setting $\mu=0$, which leads to the following equation 
\begin{eqnarray}
8r^2_+\left[e^{\frac{k}{2r_+}}(1+m^2c^2c_2)r_+-6M\right]+24kMr_+-2k^2M=0,\label{inv-eq}
\end{eqnarray}
which is independent on the coupling parameter $c_1$. Solving this equation for $r_+$ and then substituting into  Eqs. (\ref{P-eq}) and (\ref{T-eq}), one can derive the inversion curve $T_i=T_i(P_i)$. In general, it is difficult to solve exactly this equation. However, for $1+m^2c^2c_2=0$, it is easily to get exactly two solutions of Eq. (\ref{inv-eq}) as
\begin{eqnarray}
r_+=\frac{3\pm\sqrt{3}}{12}k.
\end{eqnarray}
With the positive sign, we have
\begin{eqnarray}
T^{(+)}_i&=&\frac{1}{8\pi}\left[\frac{96(2\sqrt{3}-3)M}{e^{3-\sqrt{3}}k^2}-m^2cc_1\right],\nonumber\\
P^{(+)}_i&=&\frac{27\left[96e^{\sqrt{3}}M-(2+\sqrt{3})e^3m^2cc_1k^2\right]}{2(3+\sqrt{3})^3e^3\pi k^3}.
\end{eqnarray}
For $c_1>0$, we need the following condition
\begin{equation}
\frac{M}{m^2cc_1k^2}\geq\frac{(2+\sqrt{3})e^{3-\sqrt{3}}}{96}.
\end{equation}
With the negative sign, we obtain
\begin{eqnarray}
T^{(-)}_i&=&-\frac{1}{8\pi}\left[\frac{96(2\sqrt{3}+3)M}{e^{3+\sqrt{3}}k^2}+m^2cc_1\right],\nonumber\\
P^{(-)}_i&=&\frac{27\left[96e^{-3-\sqrt{3}}M-(2-\sqrt{3})m^2cc_1k^2\right]}{2(3-\sqrt{3})^3\pi k^3},
\end{eqnarray}
which only exits if $c_1<0$ and
\begin{equation}
-\frac{M}{m^2cc_1k^2}\leq\frac{e^{3+\sqrt{3}}}{96(2\sqrt{3}+3)}.
\end{equation}
By eliminating the black hole mass $M$, we can obtain two inversion curves as
\begin{eqnarray}
T^{(+)}_i&=&\frac{\sqrt{3}+1}{6}kP^{(+)}_i+\frac{\sqrt{3}-1}{8\pi}m^2cc_1,\nonumber\\
T^{(-)}_i&=&-\frac{\sqrt{3}-1}{6}kP^{(-)}_i-\frac{\sqrt{3}+1}{8\pi}m^2cc_1.
\end{eqnarray}
First, let us consider the case $c_1>0$. In this case, it is easily to see that the curve $T^{(-)}_i\big(P^{(-)}_i\big)$ disappears. Thus, the inversion points $(P_i,T_i)$ are completely determined by the curve $T^{(+)}_i\big(P^{(+)}_i\big)$. Here, one can obtain the global minimum inversion temperature $T_{\text{min}}$ corresponding to $P^{(+)}_i=0$ as
\begin{equation}
T_{\text{min}}=\frac{\sqrt{3}-1}{8\pi}m^2cc_1.
\end{equation}
Furthermore, in this case the inversion temperature is a monotonically increasing function of the inversion pressure. And, with the inversion pressure kept fixed, the inversion temperature increases with the growth of both the non-linear parameter $k$ and the massive gravity coupling $c_1$. In the case $c_1<0$, for the large enough inversion pressure, the inversion points $(P_i,T_i)$ are determined by the curve $T^{(+)}_i\big(P^{(+)}_i\big)$. Thus, the behavior of the inversion temperature under the change of the inversion pressure, the non-linear parameter $k$ and the coupling parameter $c_1$ is like the case $c_1>0$. Whereas, for the small enough inversion pressure, the inversion points $(P_i,T_i)$ are determined by the curve $T^{(-)}_i\big(P^{(-)}_i\big)$. In contract to the large enough inversion pressure, in this case, increasing the inversion pressure, the non-linear parameter $k$ or the coupling parameter $c_1$ leads to the decreasing of the inversion temperature.

\section{\label{conclu} Conclusion}

For the black hole thermodynamics in the extended phase space at which the cosmological constant is considered as the pressure, one can introduce the concept of the heat engine for the black hole. In this paper, we consider the non-linear charged AdS black hole in the massive gravity as a heat engine. The heat cycle under consideration, given in the $P-V$ diagram, consists of two isobaric paths and two isochoric paths. It is shown that the non-linear parameter and massive gravity couplings affect significantly the efficiency of the black hole heat engine. As the non-linear parameter increases, the heat engine efficiency can increase or decrease dependently on the value region of the non-linear parameter as well as the sign of the massive gravity couplings. In addition, increasing the massive gravity couplings almost makes decreasing the heat engine efficiency.

In the extended phase space, the black hole mass is most naturally identified as the enthalpy. Thus, 
one can consider the expansion process of the black hole during which the black hole mass is kept fixed, well-known as the Joule-Thomson expansion process. We study how the presence of the non-linear elctrodynamics and the massive gravity affect the isenthalpic curves of the black hole in detail. Also, we calculate the Joule-Thomson coefficient whose sign allows to determine which of the heating or cooling will appear. Finally, we derive analytically the inversion curves, which separates the cooling region and heating region, for the case $1+m^2c^2c_2=0$.

\end{document}